\documentclass[lettersize,journal]{IEEEtran}
\usepackage{amsmath,amsfonts}
\usepackage{algorithmic}
\usepackage{array}
\usepackage[caption=false,font=normalsize,labelfont=sf,textfont=sf]{subfig}
\usepackage{textcomp}
\usepackage{stfloats}

\usepackage{url}
\usepackage{amsmath,amsthm,mathrsfs,dsfont}
\usepackage{verbatim}
\usepackage{bbm}
\usepackage{amssymb}
\usepackage{cuted}
\usepackage{graphicx}
\usepackage{tikz}
\usetikzlibrary{quantikz}
 \usepackage{algorithm}
\usepackage{algorithmic} 
\def\BibTeX{{\rm B\kern-.05em{\sc i\kern-.025em b}\kern-.08em
    T\kern-.1667em\lower.7ex\hbox{E}\kern-.125emX}}
\usepackage{balance}
\begin{document}
\title{Robust and optimal loading of general classical data into quantum computers}
\author{Xiao-Ming~Zhang
\IEEEcompsocitemizethanks{\IEEEcompsocthanksitem Xiao-Ming Zhang was with the 
Key Laboratory of Atomic and Subatomic Structure and Quantum Control (Ministry of Education), Guangdong Basic Research Center of Excellence for Structure and Fundamental Interactions of Matter, School of Physics, South China Normal University, Guangzhou 510006, China;
 Guangdong Provincial Key Laboratory of Quantum Engineering and Quantum Materials, Guangdong-Hong Kong Joint Laboratory of Quantum Matter, South China Normal University, Guangzhou 510006, China; and the Center on Frontiers of Computing Studies, School of Computer Science, Peking University, Beijing, China. 
E-mail: phyxmz@gmail.com
}
}



\maketitle

\begin{abstract}
As standard data loading processes, quantum state preparation and block-encoding are critical and necessary processes for quantum computing applications, including quantum machine learning, Hamiltonian simulation, and many others. Yet, existing protocols suffer from poor robustness under device imperfection, thus limiting their practicality for real-world applications. Here, this limitation is overcome based on a fanin process designed in a tree-like bucket-brigade architecture. It suppresses the error propagation between different branches, thus exponentially improving the robustness compared to existing depth-optimal methods. Moreover, the approach here simultaneously achieves the state-of-the-art fault-tolerant circuit depth, gate count, and STA. As an example of application, we show that for quantum simulation of geometrically local Hamiltonian, the code distance of each logic qubit can potentially be reduced exponentially using our technique. We believe that our technique can significantly enhance the power of quantum computing in the near-term and fault-tolerant regimes.
\end{abstract}


\section{introduction}
An end-to-end realization of quantum computing requires the loading of classical data to a quantum device. For example, in quantum simulation,  \textit{block-encoding}~\cite{Low.19,Chakraborty.19} is typically used for loading many-body Hamiltonians, through which the nearly-optimal dynamic simulation and ground state (energy)  estimation can be realized.  In the context of quantum machine learning, one should load the classical data, e.g. figures, language and other types of information into a quantum state. One of the standard approaches is called amplitude encoding, which is equivalent to the process of \textit{quantum state preparation}~\cite{Long.01,Grover.02,Mottonen.05,Low.24,Zhang.21,Sun.21,Rosenthal.21,Clader.22,Zhang.22,Yuan.22,Gui.23,Zhang.24}.  The study of the quantum state preparation also has its fundamental motivations, as it indicates the space-time resource required to transform one pure quantum state to another.  

Various protocols have been proposed in the literature to realize quantum state preparation. For example, Long, Sun~\cite{Long.01} and Grover, Rudolph~\cite{Grover.02} have independently proposed iterative preparation methods based on multi-controlled-rotations. Subsequent works have improved the single- and two-qubit gate count to $O(N)$, which is optimal (e.g. ~\cite{Mottonen.05,Plesch.11}). Although a large gate count is inevitable in general, it is possible to trade time (circuit depth) for space (ancillary qubit). Recently, low-depth quantum state preparation with $\Theta(n)$ circuit depth that matches the lower bound~\cite{Zhang.21,Sun.21} has been achieved by Sun \textit{et. al.}~\cite{Sun.21}, and subsequently by several other protocols~\cite{Sun.21,Rosenthal.21,Zhang.22,Yuan.22,Gui.23}, provided sufficient number of ancillary qubits. These results indicate an ultimate speed limit for loading general classical data to a quantum device. Despite the remarkable progress, current protocols are far from practical. On one hand, the robustness of~\cite{Sun.21,Rosenthal.21,Zhang.22,Yuan.22,Gui.23} cannot be guaranteed. The worst-case single- and two-qubit gate count of state preparation is $O(N)$, regardless of the space-time trade-off. A direct evaluation indicates that to achieve a constant  preparation fidelity, one should suppress the  gate error to the level of $O(N^{-1})$. For applications on large data sets, this requirement is too stringent to be practical, especially for near-term quantum devices. Even in the fault-tolerant setting, the gate error requirement of $O(N^{-1})$ is also challenging. Take the surface code~\cite{Fowler.12} scheme as an example, the code distance of each logic qubit should increase polynomially with $n$. This means that a substantial amount of classical data processing and corrections gates are required, rendering the vanishing of quantum advantages.

 On the other hand, most of the existing protocols (e.g. ~\cite{Sun.21,Rosenthal.21,Clader.22,Yuan.22,Gui.23}) assume fully connectivity, which are not friendly for current quantum devices. In superconducting circuit systems, qubits are typically connected by couplers~\cite{Arute.19,Wu.21}, and only nearest-neighbor interaction is available. There are other systems where better connectivity is available, such as trapped ion~\cite{Moses.23} and neutral-atom arrays~\cite{Bluvstein.24}. However, simultaneous rearrangement of connectivities requires complicated shuttling, which is time-costly and may substantially affect the control accuracy. Although protocols in~\cite{Zhang.22,Zhang.24} have sparse connectivity, the architecture is still far from optimal.

Besides, the bucket-brigade quantum random access memory (QRAM)~\cite{Giovannetti.08,Veras.20,Hann.21} enjoys both robustness and simple connectivity. The preliminary aim of QRAM protocols~\cite{Giovannetti.08,Veras.20,Hann.21} is to perform the specific transformation
$|j\rangle|0\rangle\rightarrow|j\rangle|D_j\rangle$ coherently for $0\leq j\leq N-1$, where  $D_j$ is binary data to be encoded, while the generalization to nonbinary $D_j$ can be realized by adding a pointer~\cite{Zhang.22}. Bucket-brigade QRAM stands out due to its provable noise resiliency~\cite{Arunachalam.15,Hann.21}. Moreover, qubits in this architecture are connected as a binary tree. Due to its simplicity, various schemes have been proposed to realize the bucket-brigade QRAM in different systems, such as neutral atom~\cite{Giovannetti.08_2,Hong.12}, superconducting circuit~\cite{Hann.19}, spin-photon network~\cite{Chen.21}, etc. 

Unfortunately, QRAM per se is only a special data loading process, which is not sufficient for many applications. The generalization of bucket-brigade mechanism to arbitrary quantum state preparation is highly nontrivial: If one performs state preparation by applying QRAM iteratively in a naive way, this will introduce a significant circuit depth overhead [increased to $O(n^2)$ as opposed to $O(n)$], thus substantially reduce both the efficiency and robustness. It remains an outstanding question for a general state preparation task, whether robustness and the optimality of circuit complexity can be achieved simultaneously.

In this work, we develop a novel fanin process to enable the bucket-brigade preparation of general quantum states.  Compared to existing depth-optimal methods~\cite{Sun.21,Rosenthal.21,Zhang.22,Yuan.22,Gui.23}, our approach overcomes both the robustness and connectivity challenges, and at the same time improves the circuit complexity. In particular, the infidelity scaling is exponentially improved from $O(N)$ to $O(\text{polylog}(N))$ under a fixed noise level. The hardware of our approach is as simple as the binary tree architecture --- each qubit connects to at most three other qubits, which is optimal. We also generalize our technique to the block-encoding of general matrices and LCU, showing similar noise-robustness and circuit complexities. As a direct consequence, we show that for the fault-tolerant simulation of geometrically local Hamiltonian, the code distance of each logic qubit can be reduced from $O(\text{polylog}(n))$ with methods in~\cite{Long.01,Grover.02,Mottonen.05,Low.24,Clader.22,Sun.21,Rosenthal.21,Zhang.22,Yuan.22,Gui.23,Zhang.24} to $O(\text{polyloglog}(n))$ with our methods.

The remaining part of the manuscript is organized as follows. In Sec.~\ref{sec:pre}, we give some introduction about the basic idea of qubit, quantum state, and its preparation. In Sec.~\ref{sec:main}, we summarize our main results. We then present the explicit implementation of the 2-qubit-per-node protocol in Sec.~\ref{sec:2}, which is relatively simple and has infidelity scaling $1-F\leq A\varepsilon n^3$.  The improved 3-qubit-per-node protocol is presented in Sec.~\ref{sec:3bit}, which has improved infidelity scaling to $1-F\leq A\varepsilon n^2$, and better circuit complexities. In Sec.~\ref{sec:be}, we  generalize our techniques to block-encoding. In Sec.~\ref{sec:cd}, we give a conclusion and further discussions.

\begin{table*}[t]
\caption{Comparison to  some typical state preparation protocols with $\tilde{O}(n)$ circuit depth. The estimation of infidelity scalings for~\cite{Sun.21,Zhang.22,Gui.23,Zhang.24} are based on direct counting of the total gate counts.  \label{tab:1}}
\begin{center}
\begin{tabular}{c|c|c|c|c|c}
\hline \cline{1-6}
Protocols &Infidelity scaling&Connectivity &Count& Depth &STA\\
\hline
Ref~\cite{Sun.21}&$O(N \varepsilon)$& all-to-all& $O(N\log(N/\varepsilon))$& $O(n\log(N/\varepsilon))$& $O(Nn\log(N/\varepsilon))$\\
Ref~\cite{Zhang.22,Zhang.24}&$O(N \varepsilon)$&degree 4&  $\boldsymbol{O\left(N\log(1/\varepsilon)\right)}$ & $O\left(n\log (n/\varepsilon) \right)$&$O(Nn\log(n/\varepsilon))$\\
Ref~\cite{Gui.23}&$O(N \varepsilon)$&all-to-all& $O\left(N\log(n/\varepsilon)\right)$& $\boldsymbol{O(n+\log(1/\varepsilon))}$&$O\left(N\log(n/\varepsilon)\right)$\\
2-qubit-per-node&$O(n^3 \varepsilon)$&\textbf{degree $3$}&  $\boldsymbol{O\left(N\log(1/\varepsilon)\right)}$ & $O\left(n\log(n/\varepsilon) \right)$& $O\left(N\log(n/\varepsilon)\right)$\\
3-qubit-per-node&$\boldsymbol{O(n^2 \varepsilon)}$ & \textbf{degree $3$}& $\boldsymbol{O\left(N\log(1/\varepsilon)\right)}$ & $\boldsymbol{O\left(n+\log(1/\varepsilon) \right)}$& $\boldsymbol{O\left(N\log(1/\varepsilon)\right)}$ \\
\hline \cline{1-6}
\end{tabular}
\end{center}
\end{table*}

\begin{figure*}[h]
    \centering
          \includegraphics[width=1.8\columnwidth]{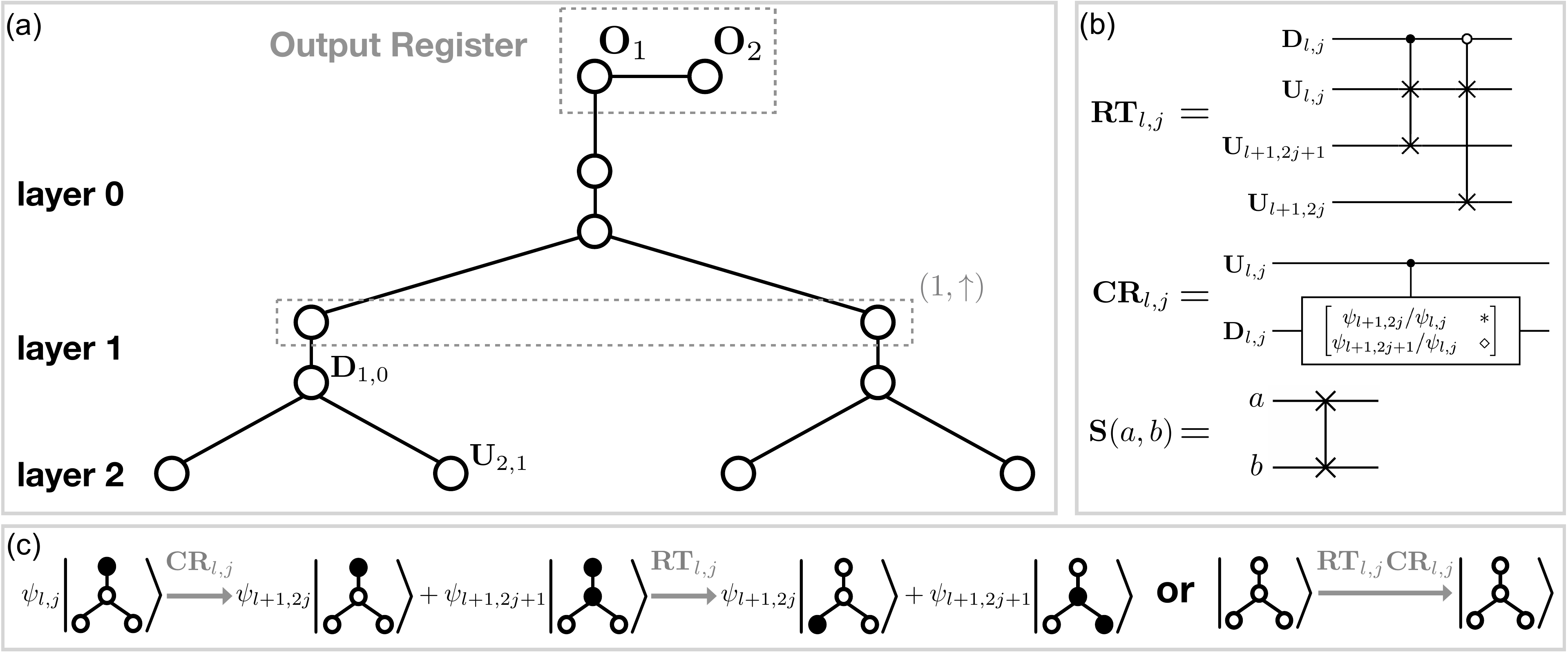}
       \caption{(a) Hardware architecture of quantum state preparation protocols and corresponding notations in the main text. We take $n=2$ as an example. Each circle represent a qubit, and each line represents the connection between a pair of qubits. (b) Definitions of routing ($\textbf{RT}_{l,j}$), controlled-rotation ($\textbf{CR}_{l,j}$), and swap $\textbf{S}(a,b)$ operations. In the operation $\textbf{CR}_{l,j}$, labels $*$ and $\diamond$ represent some values that make the matrix to be a unitary. (c) Sketch of how quantum state transforms during each operation in the fanin phase.} \label{fig:1}
\end{figure*} 

\section{Preliminaries}\label{sec:pre}
The unit of quantum computing the quantum bit, abbreviated as \textit{qubit}. It is the quantum analogue of a classical \textit{bit}. Different from classical bit that can only be in one of two states (0 or 1), a qubit can be at a superposition. Specifically, the state of a qubit is represented as a vector in a two-dimensional complex Hilbert space $\psi_{\text{qubit}}=\begin{pmatrix}\alpha\\\beta\end{pmatrix}$. In the Dirac notation, the quantum state of a qubit can be represented as $|\psi_{\text{qubit}}\rangle=\alpha|0\rangle+\beta|1\rangle$. Here, $\alpha$ and $\beta$ represents the amplitude of the state $|0\rangle$ and $|1\rangle$, which can be complex, and satisfies $|\alpha|^2+|\beta|^2=1$. 
For a system with $n$ qubits, its quantum state can be the superposition of all possible bitstrings, i.e. $|\psi\rangle=\sum_{j=0}^{N-1}\alpha_j|j\rangle$, for some $\sum_{j=0}^{N-1}|\alpha|^2=1$, where we have defined $N=2^n$, and $|j\rangle$ represents a bitstring.

In a closed system, all allowed quantum operations can be represented as unitary operators $U$ which transfer a quantum state to another in the form of $U|\psi_A\rangle=|\psi_B\rangle$. Given an $N$ dimensional normalized vector $[\psi_0,\psi_1,\cdots,\psi_{N-1}]$, we say that the unitary $U_{\text{sp}}$ prepares a target quantum state $|\psi\rangle\equiv\sum_{j=0}^{N-1}\psi_j|j\rangle$, from a trivial initial state $|0\rangle^{\otimes}$ if
\begin{align}\label{eq:sp}
U_{\text{sp}}|0\rangle^{\otimes n}\otimes |\text{anc}\rangle=|\psi\rangle\otimes |\text{anc}\rangle,
\end{align}
Here, $\otimes$ represents the Kronecker product, and $|\text{anc}\rangle$ is the quantum state of an ancillary state.   In general, $U_{\text{sp}}$ is a global operation applied at both the $n$-qubit target system and ancillary system. In practice, we should decompose it into some elementary operations that are allowed by quantum devices. These elementary operations can be single-qubit and two-qubit gates. In fault-tolerant setting, the operations are further decomposed into single-qubit Hadamard gate $H=\frac{1}{\sqrt{2}}\begin{pmatrix}1&1\\1&-1\end{pmatrix}$, T-gate $T=\begin{pmatrix}1&0\\0&e^{i\pi/4}\end{pmatrix}$, and two-qubit CNOT gate CNOT$=|0\rangle\langle0|\otimes\begin{pmatrix}1&0\\0&1\end{pmatrix}+|1\rangle\langle1|\otimes\begin{pmatrix}0&1\\1&0\end{pmatrix}$, because  error-corrected implementation of these operations are possible with surface code~\cite{Fowler.12}.

\section{Main results}\label{sec:main}
We have developed two protocols for quantum state preparation. Both of the protocols have the simplest connectivity, i.e. each qubit connects to at most three of other qubits, and achieves the best-known gate count $O(N\log(1/\varepsilon))$.  Under depolarization channels applied at all qubits, the state preparation infidelity scales as $O(n^3\varepsilon)$ for the 2-qubit-per-node protocol, and scales as $O(n^2\varepsilon)$ for the 3-qubit-per-node protocol.  Their circuit depths are $O(n\log(n/\varepsilon))$ and $O(n\log(1/\varepsilon))$ respectively. We also optimize the space-time-allocation (STA)---the total
time that each individual qubit must be active. The STA for two protocols are $O(N\log(n/\varepsilon))$ and $O(N\log(1/\varepsilon))$ respectively. Our main results and comparison to existing protocols are  summarized in Table.~\ref{tab:1}.

\section{2-qubit-per-node protocol}\label{sec:2}
\subsection{Hardware architecture}

As shown in Fig.~\ref{fig:1}, our 2-qubit-per-node protocol contains a bucket-brigade QRAM and an $n$-qubits output register. 
The bucket-brigade QRAM resembles an $(n+1)$ layer binary tree and each node of the tree corresponds to two qubits. To be specific, the $l$th ($0\leq l\leq n$) layer contains an upper and a lower sublayer, denoted as $(l,\uparrow)$ and $(l,\downarrow)$ respectively. An exception is that the leaf layer has only upper sublayers. Each sublayer contains totally $2^l$ qubits. We denote the $j$th qubit of $(l,\uparrow)$ and $(l,\uparrow)$ as $\textbf{U}_{(l,j)}$ and $\textbf{D}_{(l,j)}$. 
In QRAM, each qubit connects only to their parent or children. $\textbf{U}_{l,j}$ has one child $\textbf{D}_{l,j}$, and $\textbf{D}_{l,j}$ for $l\neq n$ has two children $\textbf{U}_{l+1,2j}$ and $\textbf{U}_{l+1,2j+1}$. 
 The output register contains $n$ qubits, each denoted as $\textbf{O}_{j}$ (from $j=1$ to $j=n$). They are arranged as a line with nearest-neighbor coupling, and $\textbf{O}_1$ also connect to the root of QRAM, i.e. $\textbf{U}_{0,0}$. In this architecture, each qubit connects to at most $3$ of the other qubits, which is optimal. This is  because a graph of degree $2$ can only form trivial lines or rings, qubit connections in such ways are insufficient for achieving subexponential circuit depth. 
 
\subsection{Geometrically non-local gate}
\begin{figure}[t]
    \centering
          \includegraphics[width=.9\columnwidth]{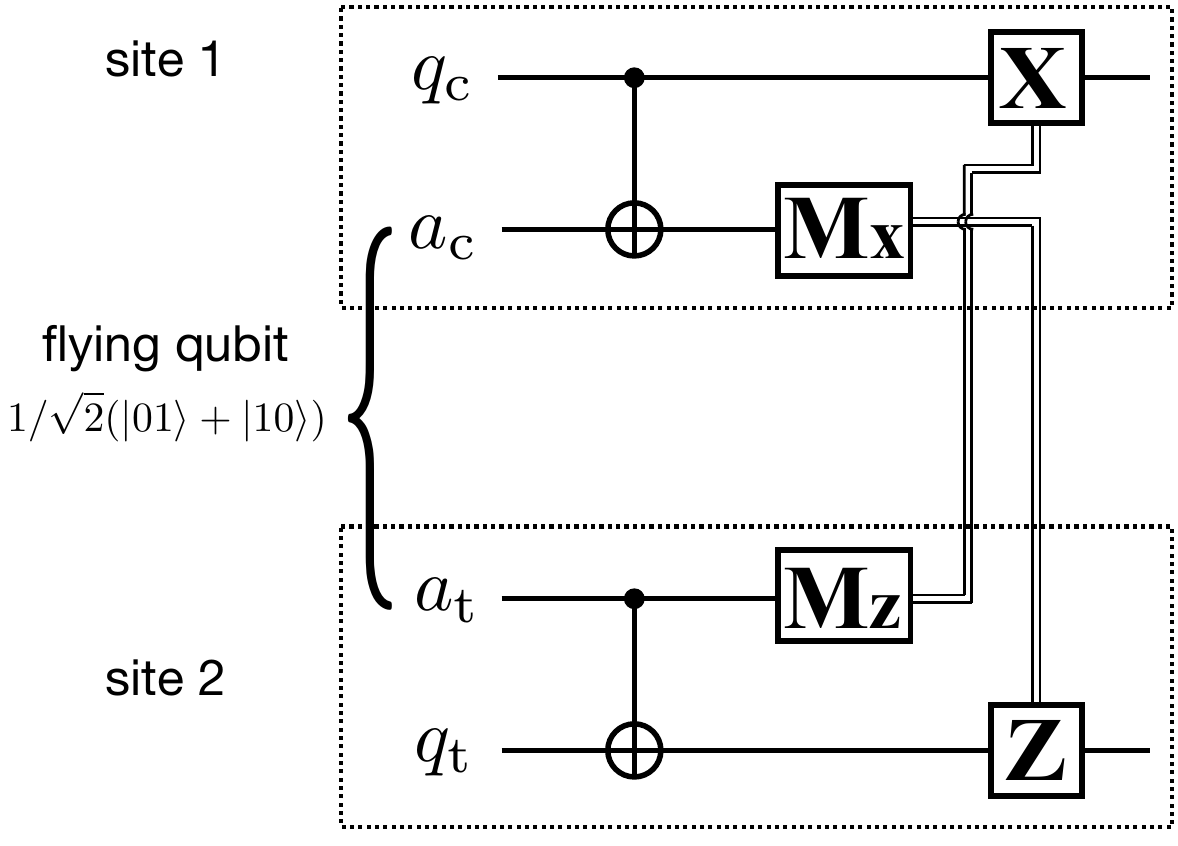}
       \caption{Teleportated CNOT gate assisted with flying qubits at Bell state.} \label{fig:lr}
\end{figure}

Similar to existing methods~\cite{Sun.21,Rosenthal.21,Zhang.22,Yuan.22,Gui.23,Zhang.24}, the geometrically long-range interaction is fundamentally inevitable. In practice, it can be realized by,  teleportated quantum gate assisted with flying qubits (e.g. photons)~\cite{Gottesman.99,Chou.18}. As shown in Fig.~\ref{fig:lr}, our goal is to implement CNOT gate with atom qubits $q_\text{c}$ and $q_\text{t}$ as the controlled and target qubits, and they are at site 1 and 2 respectively.  First, we generate a pair of entanglement photons at Bell state $1/\sqrt{2}(|01\rangle+|10\rangle)$, and denote the flying qubits as $a_\text{c}$ and $a_\text{q}$ respectively. We sent $a_\text{c}$ to site 1 and sent $a_\text{t}$ to site 2. Second,  we implement local CNOT gate at site 1 with $q_\text{c}$ and $a_\text{c}$ as controlled and target qubits. At site $2$, we implement local CNOT gate with $a_\text{t}$ and $q_\text{t}$ as controlled and target qubits. This process can be realized by spin-photon interactions. Third, we measure flying qubit $a_\text{c}$ at basis $\{|+\rangle,|-\rangle\}$. Conditioned on the measurement outcome to be $|+\rangle$, we apply $Z$ gate at qubit $q_\text{t}$. Finally, measure flying qubit $a_\text{t}$ at basis $\{|0\rangle,|1\rangle\}$, and conditioned on the measurement outcome to be $|0\rangle$, we apply $X$ gate at qubit $q_\text{c}$. It can be verified that this process is equivalent of performing non-local CNOT gate with $q_{\text{c}}$ and $q_{\text{b}}$ be the controlled and target qubits. 

Alternatively, the non-local gate can also be realized by shuttling~\cite{Moses.23,Bluvstein.24}. Also, nearest-neighbour-coupling-based implementation with fault-tolerance is also possible, at the cost of a mild extra overhead of time and space~\cite{Zhang.22}. 
 
 \begin{figure*}[!t]
          \includegraphics[width=2\columnwidth]{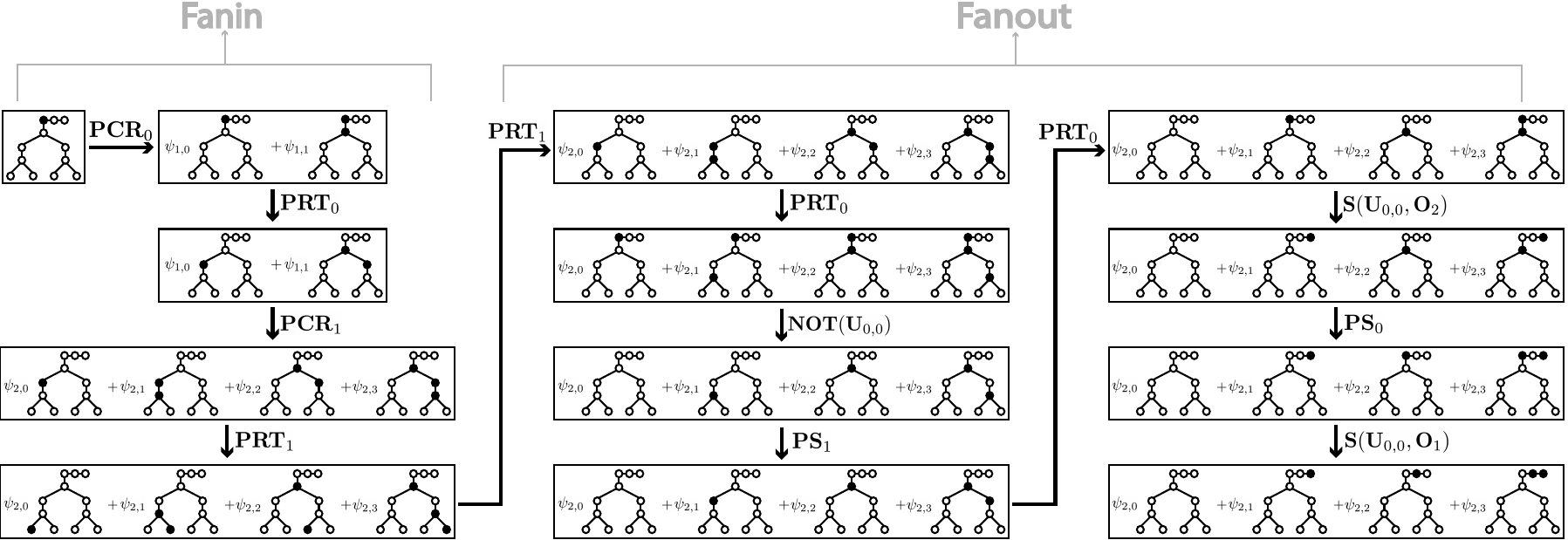}
       \caption{Sketch of the 2-qubit-per-node protocol for $n=2$ case. Hollow and solid circles represent qubits at quantum states $|0\rangle$ and $|1\rangle$ respectively. } \label{fig:skt2}
\end{figure*}

 \subsection{Fanin phase}

  In fanin phase, we only perform operations in QRAM. We begin with some notations of quantum states. Suppose $\mathcal{S}$ is a set of qubits, we use the ``activation'' representation $|\mathcal{S}\rangle$ to represent that  all qubits in $S$ are activated (i.e. at state $|1\rangle$), while all other qubits are at state $|0\rangle$. Formally, we have $|\mathcal{S}\rangle\equiv\otimes_{v\in\mathcal{V}^{\text{QRAM}}}|v\in \mathcal{S}\rangle_v$, where  $|\cdot\rangle_v$ represents the state of qubit $v$, and the ``True'' or ``False'' result of $v\in \mathcal{S}$ correspond to the binary $1$ or $0$. $\mathcal{V}^{\text{QRAM}}$ represents all qubits in QRAM.

Let $|\psi_0\rangle=|\{\textbf{U}_{0,0}\}\rangle$ be the initial state (i.e. only the root of QRAM is activated), we perform the following transformation 
\begin{align}\label{eq:enc}
|\psi_{l}\rangle\overset{}{\longrightarrow}|\psi_{l+1}\rangle,\quad |\psi_l\rangle\equiv\sum_{j=0}^{2^l-1}\psi_{l,j}|\mathscr{B}_{l,j}\rangle,
\end{align}
iteratively from $l=0$ to $l=n-1$,
where $\psi_{l,j}$ will be defined later, and $\mathscr{B}_{l,j}$ is a set of qubits that will be clarified as follows. For qubit $\textbf{D}_{(l,j)}$ at lower sublayers, we let  $\mathcal{P}_{\downarrow}\left[\textbf{D}_{(l,j)}\right]=\textbf{D}_{(l-1,\lceil j/2\rceil)}$ be its grandparent, which is also at the lower sublayers. Accordingly, we represent all ancestors of $\textbf{D}_{(l,j)}$ in the lower sublayers as $\mathscr{A}_{l,\downarrow,j}=\left\{{\mathcal{P}}_{\downarrow}^{\odot m}[\textbf{D}_{(l,j)}]\big|1\leq m\leq l \right\}$, which contains totally $l$ qubits. $|\mathscr{B}_{l,j}\rangle$ represents the following quantum state: at the subset of upper sublayers, only single qubit, $\textbf{U}_{l,j}$, is activated and it serves as a \textit{pointer}. At the subset of lower sublayers, $\mathscr{A}_{l,\downarrow,j}$ (all ancestors of $\textbf{U}_{l,j}$ at lower sublayers) is at computational basis $|j_1j_2\cdots j_l\rangle$, i.e. the first $l$ bits of $j$. All other qubits are at state $|0\rangle$. The formal definition of $\mathscr{B}_{l,j}$ is 

\begin{subequations}\label{eq:BB}
\begin{align}
\mathscr{B}'_{l,j}&=\left\{ \textbf{D}_{l',j'}\in\mathscr{A}_{l,\downarrow,j}\big|{j'}_{l'}=1\right\}\\
\mathscr{B}_{l,j}&=\mathscr{B}'_{l,j}\cup\{\textbf{U}_{l,j}\}.
\end{align}
\end{subequations}
which clarifies Eq.~\eqref{eq:enc}.

We then define $\psi_{l,j}$. Firstly, each amplitude may be represented as $\psi_{j}\equiv a_{j}\angle \phi_j$, where $a_j$ and $\phi_j$ are absolute value and argument of $\psi_j$ respectively. We set $\phi_0=0$ without loss of generality. Let $\psi_{n,j}\equiv \psi_j$, we recursively define $\psi_{l,j}=e^{i \phi_{l+1,2j}}\sqrt{a_{l+1,2j}^2+a_{l+1,2j+1}^2}$. 

We then turn to the gates required for this phase. Let $\textbf{CR}_{l,j}$ be a controlled-rotation with $\textbf{U}_{l,j}$ and $\textbf{D}_{l,j}$ as controlled and target qubits, which satisfies  (see also Fig.~\ref{fig:1}(b))
\begin{subequations}
\begin{align}
\textbf{CR}_{l,j}|1\rangle\otimes(\psi_{l,j}|0\rangle)&=|1\rangle\otimes(\psi_{l,2j}|0\rangle+\psi_{l,2j+1}|1\rangle)\\
\textbf{CR}_{l,j}|0\rangle\otimes|0\rangle&=|0\rangle\otimes|0\rangle.
\end{align}
\end{subequations}
Let $\textbf{PCR}_l=\prod_{j=0}^{2^l-1}\textbf{CR}_{l,j}$ be the parallel controlled-rotation, which can be realized with one layer of quantum circuit.  This parallel rotation is crucial for data encoding.  

Another critical operation is \textit{routing}. Let $\textbf{S}(a,b)$ be the swap gate between qubit $a$ and $b$, routing is the following transformation

\begin{align}\label{eq:rt}
|0\rangle_{\text{rt}}\langle0|\otimes\textbf{S}(\text{in},\text{lo})+|1\rangle_{\text{rt}}\langle1|\otimes\textbf{S}(\text{in},\text{ro})
\end{align}
If the routing qubit (rt) is at state $|0\rangle$, we swap the states of incident qubit (in) and left output qubit (lo); if the routing qubit is at state $|1\rangle$, we swap the states of input qubit and right output qubit (ro). We denote $\textbf{RT}_{l,j}$ as the routing operation defined in Eq.~\eqref{eq:rt} with $\textbf{U}_{l,j}$, $\textbf{D}_{l,j}$, $\textbf{U}_{l,2j}$ and $\textbf{U}_{l+1,2j+1}$ as the input, routing, left output and right output qubits respectively (see also Fig.~\ref{fig:1}(b)). Note that $\textbf{RT}_{l,j}$ for different $j$ can be implemented in parallel. Accordingly, we define  $\textbf{PRT}_l=\prod_{j=0}^{2^l-1}\textbf{RT}_{l,j}$. This parallel routing can be implemented with constant circuit depth.

With elementary gates being explained, we are ready to discuss the transformation in Eq.~\eqref{eq:enc}. We first apply parallel controlled rotation $\textbf{PCR}_l$. Except for qubits connected to the pointer (currently at $\textbf{U}_{l,j}$), other qubits at sublayers $(l,\downarrow)$ are not activated. So it can be verified that
$\textbf{PCR}_l(\psi_{l,j}|\mathscr{B}_{l,j}\rangle)\notag=\psi_{l+1,2j}|\{\mathscr{B}_{l,j}\}\rangle+\psi_{l+1,2j+1}|\mathscr{B}_{l,j}\cup\{\textbf{D}_{l,j}\}\rangle$. Then, we move the pointer from the $l$th to the $(l+1)$th layer using parallel routing operation $\textbf{PRT}_l$. Recall that $\textbf{D}_{l,j}$ are controlled qubits of our routing operations. According to the property defined by Eq.~\eqref{eq:rt}, if $\textbf{D}_{l,j}$ is not activated, the pointer moves to $\textbf{D}_{l+1,2j}$, otherwise the pointer moves to $\textbf{D}_{l+1,2j+1}$. Following the definition of $\mathscr{B}_{l,j}$, we have 
 
 \begin{subequations}
\begin{align}
\textbf{PRT}_l|\mathscr{B}_{l,j}\rangle&=|\mathscr{B}_{l+1,2j}\rangle\\
\textbf{PRT}_l|\mathscr{B}_{l,j}\cup\{\textbf{D}_{l,j}\}\rangle&=|\mathscr{B}_{l+1,2j+1}\rangle
\end{align}
\end{subequations}
Combining with the recursive definition of $\psi_{l,j}$, we have $\textbf{PCR}_l\textbf{PRT}_l|\psi_l\rangle=|\psi_{l+1}\rangle$. Therefore, at the $l$th step, it suffices to implement $\textbf{PCR}_{l}\textbf{PRT}_{l}$ to realize the transformation in Eq.~\eqref{eq:enc}. A sketch about how quantum state transforms during the $\textbf{RT}_{l,j}$ and $\textbf{CR}_{l,j}$ is illustrated in Fig.~\ref{fig:1}(c). A sketch of the complete fanin process is also illustrated in Appendix.

\subsection{Fanout stage}
In this stage, our goal is to prepare the output register to the quantum state in Eq.~\eqref{eq:sp}, while uncomputing the QRAM. In other words, we perform the basis transformation $|\mathscr{B}_{n,j}\rangle|0\cdots0\rangle_{\text{out}}\rightarrow|\varnothing\rangle|j\rangle_{\text{out}}$, where $|\cdot\rangle_{\text{out}}$ is the quantum state of output register in binary representation, while the state of QRAM is still in activation representation. This transformation has been introduced in~\cite{Hann.21} for binary data, and has subsequently been generalized to continuous data by adding an extra pointer~\cite{Zhang.22}.  

We define the shorthand  $\textbf{PRT}_{a:b}\equiv\textbf{PRT}_a\cdots\textbf{PRT}_{b+1}\textbf{PRT}_b$ for some $b>a$. We first perform operation $\textbf{PRT}_{0:n-1}$. The pointer is then moved to the root of the QRAM, i.e.
$\textbf{PRT}_{0:n-1}|\mathscr{B}_{n,j}\rangle=|\mathscr{A}_{n,\downarrow,j}\cup\{\textbf{U}_{0,0}\}\rangle$ for arbitrary $j$.
So we can then apply $\textbf{NOT}(\textbf{U}_{0,1})$ (i.e. NOT gate at qubit $\textbf{U}_{0,0}$) to uncompute the pointer. The basis is then transferred to $|\mathscr{B}_{n,j}'\rangle|0\cdots0\rangle_{\text{out}}$.

We then define $|\Psi_{l,j}\rangle=|\mathscr{B}_{l,j_{1:l}}'\rangle\otimes|0\cdots 0 j_{l+1}\cdots j_n\rangle_{\text{out}}$. The current basis and target basis correspond to $l=n$ and $l=0$ respectively. We will then perform the basis transformation $|\Psi_{l+1,j}\rangle\rightarrow|\Psi_{l,j}\rangle$ iteratively.  We define $\textbf{PS}_l=\prod_{j=1}^{2^l}\textbf{S}(\textbf{U}_{l,j},\textbf{D}_{l,j})$ as the parallel swap gate applied between sublayers $(l,\uparrow)$ and $(l,\downarrow)$. By applying $\textbf{PS}_l$ to  $|\Psi_{l+1}\rangle$, activations at sublayer $(l,\downarrow)$ are transferred to sublayers $(l,\uparrow)$.
 We then implement routing $\textbf{PRT}_{l-1:0}$, after which the sublayer $(l,\uparrow)$ is uncomputed, while the root of QRAM is prepared at state $|j_{n-l}\rangle$. Therefore, by further performing swap gate $\textbf{S}(\textbf{U}_{0,0}, \textbf{O}_{l+1})$, we complete the transformation. Note that $\textbf{S}(\textbf{U}_{0,0}, \textbf{O}_{l+1})$ is a non-local operation, which should be decomposed into totally $(l+1)$ steps of local swap gates applied at pairs of connected qubits. To conclude, let $\textbf{Fanout}(l)\equiv\textbf{S}(\textbf{U}_{0,1},\textbf{O}_{l+1})\textbf{PRT}_{l-1:0}\textbf{PS}_{l}$ (see Algorithm.~\ref{alg:fan}),  we have  
\begin{align}\label{eq:fanone}
\textbf{Fanout}(l)|\Psi_{l+1}\rangle=|\Psi_{l}\rangle
\end{align}
for $0\leqslant l\leqslant n-1$. Transformation $\textbf{Fanout}(l)$ has circuit depth $O(l)$. If we naively implement Eq.~\eqref{eq:fanone} for different $l$ sequentially, the total circuit depth is $O(n^2)$. Fortunately, we have a more efficient way. We can start $\textbf{Fanout}(l)$, idle for three steps, and then start $\textbf{Fanout}(l-1)$. In this way, operations $\textbf{Fanout}(l)$ and $\textbf{Fanout}(l-1)$ will still not affect each other. The pseudo code of fanout process begins at the third line of Algorithm.~\ref{alg:sp}. See also Fig.~\ref{fig:skt2} for illustration.  The fanout, and also the  entire state preparation process, has circuit depth $O(n)$. 

 \begin{algorithm} [t]
\caption{2-qubit-per-node quantum state preparation} 
\label{alg:sp}  
\begin{algorithmic}
\STATE \textbf{for} $l=0,\cdots,n-1$:
\STATE \quad implement  $\textbf{PRT}_{l}\textbf{PCR}_{l}$
\STATE \textbf{for} $m=0$ to $n$:
\STATE \quad start $\textbf{Fanout}(n-m)$ 
\STATE \quad idle for $3$ steps
\end{algorithmic} 
\end{algorithm} 
\begin{algorithm} [t]
\caption{ Subroutine \textbf{Fanout}($l$) for 2-qubit-per-node quantum state preparation}  
\label{alg:fan}  
\begin{algorithmic}
\STATE \textbf{if} $l\neq n$, implement $\textbf{PS}_{l}$   \quad\quad\quad\quad\quad\quad\quad$\#$ takes $1$ step
\STATE implement $\textbf{PRT}_{l-1:0}$         \,\quad\quad\quad\quad\quad\quad\quad\quad\;$\#$ takes $l$ steps
\STATE \textbf{if} $l\neq n$, implement $\textbf{S}(\textbf{U}_{0,0},\textbf{O}_{l+1})$ \quad\quad\;\;\;$\#$ takes $l$ steps
\STATE \textbf{if} $l=n$, $\textbf{NOT}(\textbf{U}_{0,0})$ \;\quad\quad\quad\quad\quad\quad\quad\quad$\#$ takes $l$ steps
\end{algorithmic} 
\end{algorithm} 

\subsection{Robustness}
One of the crucial advantages of bucket-brigade architecture is noise resiliency. 
In ~Appendix.~\ref{sec:nr},
we show that the error of our scheme scales only polylogarithmically with $n$. In particular, we consider the local depolarization model that is standard in the noisy quantum circuit study~\cite{Boixo.18,Bravyi.20}, although the results in this work are expected to be also valid for more general scenarios. The specific model is as follows, after each layer of the elementary single- and two-qubit gates, depolarization channel 
\begin{align}
(1-\varepsilon)\mathcal{I}+\varepsilon/3(\mathcal{X}+\mathcal{Y}+\mathcal{Z})
\end{align}
 is applied on \textit{all} qubits with fixed $\varepsilon$, where $\mathcal{X}$, $\mathcal{Y}$, $\mathcal{Z}$ and $\mathcal{I}$ are single qubit Pauli $X$, $Y$, $Z$ and $I$ channels respectively. Under local Pauli noise, the state preparation infidelity for Algorithm.1 satisfies $1-F\leqslant A\varepsilon n^3$ 
for some constant $A$. As a comparison, for a general quantum circuit with $O(2^n)$ elementary gates, the total infidelity scales exponentially with $n$.

The main idea of our proof about noise robustness is as follows. The noisy circuit can be decomposed into the linear combination of unitary evolutions, and each unitary evolution represents a specific space-time error configuration $c$. By a careful analysis on how error propagates between different branches of the QRAM, the final output state can be expressed as $|\tilde{\psi}(c)\rangle_{\text{out}}=\sum_{j\in g'(c)}\psi_{j}|f(c)\rangle_{\text{qram}}\otimes|j\rangle_{\text{out}}+|\text{garb}\rangle$. Where $g'(c)$ represents some \textit{error-free} branches that error will never propagate into it and $|\text{garb}\rangle$ is an unnormalized garbage state orthogonal to the first term. An important fact is that after tracing out QRAM part of $|\tilde{\psi}(c)\rangle_{\text{out}}$, the infidelity satisfies $1-F(c)\leqslant\sum_{j\in g'(c)}|\psi_{j}|^2\equiv\Lambda'(c)$. In sampling different error configuration $c$, we have $\mathbb{E}[\Lambda'(c)]\geqslant(1-A\varepsilon n^3)$. The cubic infidelity scaling then follows from the concavity of fidelity.

\section{3-qubit-per-node protocol}\label{sec:3bit}
The Clifford$+T$ complexity, which is important for fault-tolerant implementation, is not yet optimal for the protocol above. 
In this architecture, a middle sublayer is inserted between $(l,\uparrow)$ and  $(l,\downarrow)$, while each qubit is still connected to at most 3 other qubits. One of the advantages is that we can use the pre-rotation~\cite{Gui.23} technique, i.e. rotations encoding amplitudes $\psi_{l,j}$ are implemented prior to the routing operations. This allows us to simultaneously achieve the linear Clifford$+T$ circuit depth, gate count number  and STA (see Tabel.~\ref{tab:1}). More importantly, the 3-qubit-per-node protocol can further improve the noise robustness. All routing operations should be controlled by extra pointer qubits in the middle sublayers. This revision can  block all the error propagation from bad branches to good branches, and hence improve the infidelity to 
\begin{align}\label{eq:F2}
1-F\leqslant A\varepsilon n^2.
\end{align}
See Appendix.~\ref{sec:3rb} for details.

It is worth noting that a similar idea is also applicable to improve the robustness of qubit-based QRAM. More specifically, it is known that \textit{qutrit}-based QRAMs have quadratic infidelity scaling~\cite{Hann.21}. By replacing the qutrits by the combination of two qubits, one can improve the infidelity scaling from cubic to quadratic. Yet, our protocol for 
Eq.~\eqref{eq:sp} is more than this replacement, because the pre-rotation technique~\cite{Gui.23} enables the further improvement of Clifford+$T$ complexities.

\subsection{Hardware architecture and basic operations}
In our 3-qubit-per-node protocol, each layer contains $3$ sublayers. The upper, middle, and lower sublayers of the $l$th layer are denoted as  $(l,\uparrow)$, $(l,\bullet)$, and $(l,\downarrow)$ respectively. Each sublayer contain $2^l$ qubits, each denoted as $\textbf{U}_{l,j}$, $\textbf{M}_{l,j}$, $\textbf{D}_{l,j}$ respectively with $0\leqslant j\leqslant 2^l-1$. The children of $\textbf{U}_{l,j}$, $\textbf{M}_{l,j}$ and $\textbf{D}_{l,j}$ are $\{\textbf{M}_{l,j}\}$, $\{\textbf{D}_{l,j}\}$, and $\{\textbf{U}_{l+1,2j},\textbf{U}_{l+1,2j+1}\}$ respectively. Moreover, the output register is identical to the 2-qubit-per-node protocol. The hardware architecture contains more qubits (totally $6N-3$ qubits), but each qubit is still connected to at most 3 other qubits. 

We then introduce some basic operations. Let $r_{l,j}\equiv\begin{pmatrix}
\psi_{l+1,2j}/\psi_{l,j}&*\\
\psi_{l+1,2j+1}/\psi_{l,j}&\diamond\\
\end{pmatrix}$, where $*$ and $\diamond$ are some complex values that make $r_{l,j}$ be a unitary. We have the following basic operations. 

\begin{itemize}
\item $\textbf{R}_{l,j}$: rotation $r_{l,j}$ applied at qubit $\textbf{D}_{l,j}$
\item $\overline{\bold{CR}}_{l,j}$  controlled rotation $|0\rangle\langle0|\otimes r_{l,j}^\dag+|1\rangle\langle1|\otimes\mathbb{I}$ with $\textbf{M}_{l,j}$ and $\textbf{D}_{l,j}$ as controlled and target qubits
\item $\textbf{CNOT}_{l,j}$: CNOT gate with $\textbf{U}_{l,j}$ and $\textbf{M}_{l,j}$ be the control and target qubits
\item $\textbf{CRT}_{l,j}$: Five-qubit-gate 
\begin{align}
&|0\rangle_{\textbf{M}_{l,j}}\langle0|\otimes\mathbb{I}\notag\\
+&|1\rangle_{\textbf{M}_{l,j}}\langle1|\otimes|0\rangle_{\textbf{D}_{l,j}}\langle0|\otimes \textbf{S}(\textbf{U}_{l,j},\textbf{U}_{l+1,2j})\notag\\
+&|1\rangle_{\textbf{M}_{l,j}}\langle1|\otimes|1\rangle_{\textbf{D}_{l,j}}\langle1|\otimes \textbf{S}(\textbf{U}_{l,j},\textbf{U}_{l+1,2j+1})\notag
\end{align}
\item $\textbf{S}_{l,j}^{(\uparrow,\bullet)}$: swap gate between $\textbf{U}_{l,j}, \textbf{M}_{l,j}$
\item  $\textbf{S}_{l,j}^{(\uparrow,\downarrow)}$: swap gate between $\textbf{U}_{l,j}, \textbf{D}_{l,j}$
\end{itemize}
Accordingly, we define the following parallel operations
\begin{align}
\textbf{ENCODE}\equiv\sum_{l=0}^{n-1}\sum_{j=0}^{2^l-1}\textbf{R}_{l,j},\\
\textbf{DECODE}\equiv\sum_{l=0}^{n-1}\sum_{j=0}^{2^l-1}\overline{\textbf{CR}}_{l,j},
\end{align}
which encode or decode the rotation angles. We also define
\begin{align}
\textbf{PCNOT}_l\equiv\sum_{j=0}^{2^l-1}\textbf{R}_{l,j},\quad
\textbf{PCRT}_l\equiv\sum_{j=0}^{2^l-1}\textbf{R}_{l,j},\\
\textbf{PS}_{l}^{(\uparrow,\bullet)}\equiv \sum_{j=0}^{2^l-1}\textbf{S}_{l,j}^{(\uparrow,\bullet)},\quad
\textbf{PS}_{l}^{(\uparrow,\downarrow)}\equiv \sum_{j=0}^{2^l-1}\textbf{S}_{l,j}^{(\uparrow,\downarrow)}
\end{align}
that act on a specific layer $0\leqslant l\leqslant n$. All parallel operations above can be implemented with $O(1)$ layer of single- and two-qubit gates.

\begin{algorithm} [t]
\caption{3-qubit-per-node quantum state preparation}
\label{alg:sp3}  
\begin{algorithmic}
\STATE implement $\textbf{PR}$
\STATE \textbf{for} $l=0,\cdots,n-1$:
\STATE \quad implement  $\textbf{PCNOT}_{l}$
\STATE \quad implement  $\textbf{PCRT}_{l}$
\STATE implement $\textbf{P}\overline{\bold{C}\textbf{R}}$
\STATE \textbf{for} $m=0$ to $n$:
\STATE \quad start $\textbf{Fanout}(n-m)$ 
\STATE \quad idle for $6$ steps
\end{algorithmic} 
\end{algorithm} 

\begin{algorithm} [t]
\caption{ Subroutine $\textbf{Fanout}(l)$ for 3-qubit-per-node quantum state preparation}  
\label{alg:fan3}  
\begin{algorithmic}
\STATE  \textbf{if} $l= n$:
\STATE\quad \textbf{for} $l'=1$ to $l'=l$ 
\STATE\quad \quad implement $\textbf{PRT}_{l-l'}$         \,\quad\quad\quad\quad\quad\quad\quad\quad$\#$ takes $1$ steps
\STATE  \textbf{else if} $l\neq n$:
\STATE\quad implement $\textbf{PS}_{l}^{(\uparrow,\bullet)}$   \quad\quad\quad\quad\quad\quad\quad\quad\quad\;\;$\#$ takes $1$ step
\STATE\quad implement $\textbf{PRT}_{l-2}\textbf{PRT}_{l-1}$         \,\quad\quad\quad\quad\quad\quad$\#$ takes $2$ steps
\STATE\quad implement $\textbf{PS}_{l}^{(\uparrow,\downarrow)}$   \quad\quad\quad\quad\quad\quad\quad\quad\quad\;\;$\#$ takes $2$ step
\STATE\quad \textbf{for} $l'=1$ to $l'=l-2$ 
\STATE\quad \quad implement $\textbf{PRT}_{l-l'}\textbf{PRT}_{l-l'-2}$       \quad\quad\quad\quad$\#$ takes $1$ steps
\STATE\quad $\textbf{NOT}(\textbf{U}_{0,1})$ \;\quad\quad\quad\quad\quad\quad\quad\quad\quad\quad\quad\quad$\#$ takes $1$ steps
\STATE\quad  implement $\textbf{PRT}_{0}\textbf{PRT}_{1}$         \quad\quad\quad\quad\quad\quad\quad\;\;$\#$ takes $2$ steps
\STATE\quad $\textbf{S}(\textbf{U}_{0,1},\textbf{O}_l)$ \;\quad\quad\quad\quad\quad\quad\quad\quad\quad\quad\quad\quad$\#$ takes $l$ steps
\end{algorithmic} 
\end{algorithm} 


 \begin{figure*}[!t]
 \centering
          \includegraphics[width=1.8\columnwidth]{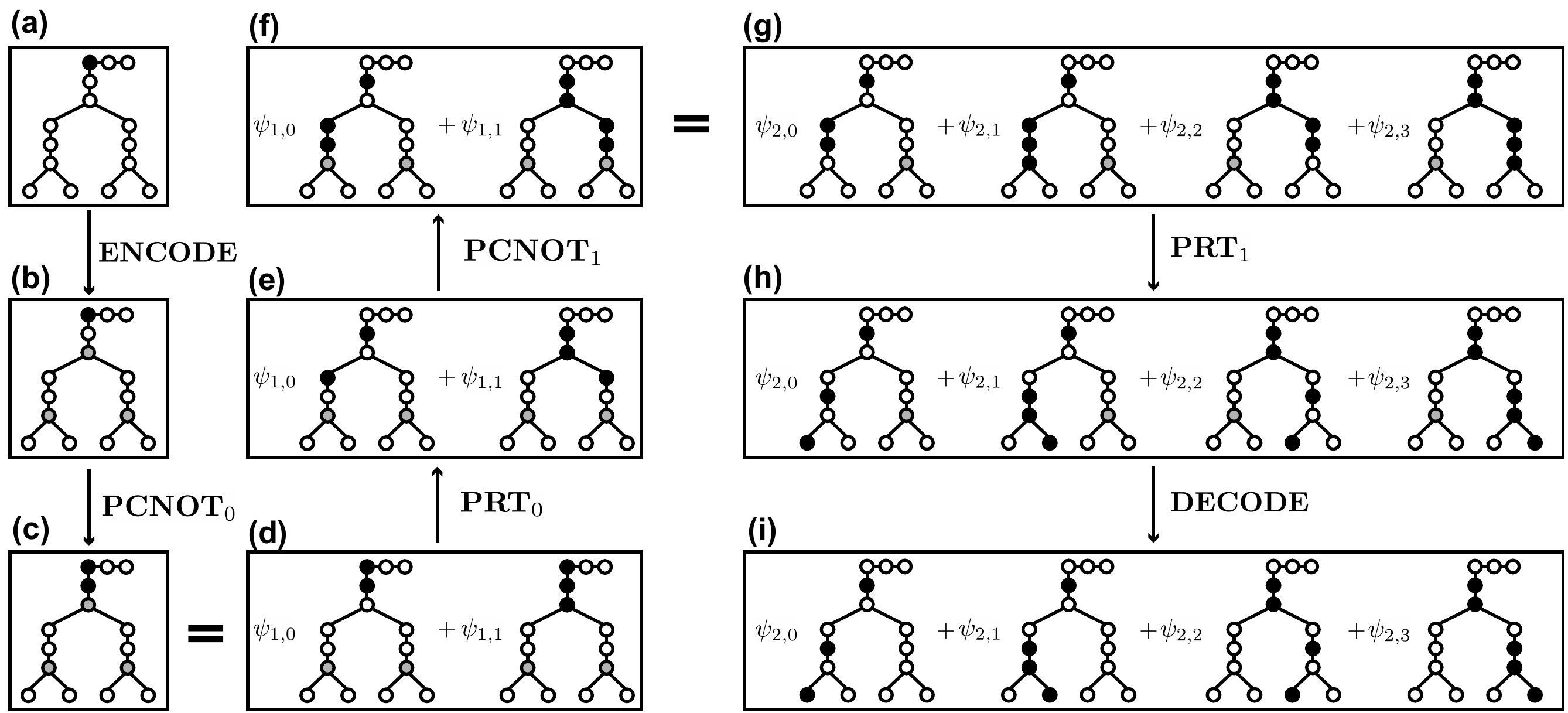}
       \caption{Sketch of the fanin process of 3-qubit-per-node protocol for $n=2$ case. Hollow and solid circles represent qubits at quantum state $|0\rangle$ and $|1\rangle$ respectively.} \label{fig:skt3in}
\end{figure*}

\subsection{Fanin phase}
The pseudo-code of our quantum state preparation algorithm is illustrated in Algorithm.~\ref{alg:sp3} and Algorithm.~\ref{alg:fan3}. The fanin process corresponds to line 1-5 in Algorithm.~\ref{alg:sp3} . An example for $n=2$ is also illustrated in Fig.~\ref{fig:skt3in}. Our method is inspired by the pre-rotation technique in~\cite{Gui.23},  which encodes angles $\{\psi_{l,j}\}$ before the controlled routing. The advantage is that pre-rotation can push the Clifford$+T$ depth to a linear scaling. For clarity, we discuss the single- and two-qubit decomposition in this section, while the Clifford$+T$ decomposition will be introduced in Sec.~\ref{sec:ct}. 

Let  $\mathscr{D}=\{\textbf{D}_{l,j}|0\leqslant l\leqslant n-1, 0\leqslant j\leqslant 2^l-1\}$ be the set of all qubits in the lower sublayers. We first implement parallel rotation $\textbf{ENCODE}$, and $\mathscr{D}$ is prepared as (Fig.~\ref{fig:skt3in} (a)-(b))

\begin{align}\label{eq:mscrd}
|\theta\rangle_{\mathscr{D}}\equiv\bigotimes_{\textbf{D}_{l,j}\in\mathscr{D}}\left(\psi_{l+1,2j}/\psi_{l,j}|0\rangle_{\textbf{D}_{l,j}}+\psi_{l+1,2j+1}/\psi_{l,j}|1\rangle_{\textbf{D}_{l,j}}\right).
\end{align}
Qubits in $\mathscr{D}$ serve as the routing qubits of our subsequent controlled-routing operations. 

 Let $\mathscr{A}_{l,\downarrow,j}$  be all ancestors of $\textbf{D}_{l,j}$ at lower sublayers, and we further define $\mathscr{D}_{l,j}=\mathscr{D}-\mathscr{A}_{l,\downarrow,j}$. $\mathscr{D}_{l,j}$ represents all routing qubits in $\mathscr{D}$ that are irrelevant to the operation $\textbf{CRT}_{l,j}$ during our fanin process. We also define 
 
\begin{align}
&|\theta\rangle_{\mathscr{D}_{l,j}}\notag\\
\equiv&\bigotimes_{\textbf{D}_{l',j'}\in\mathscr{D}_{l,j}}\left(\frac{\psi_{l'+1,2j'}}{\psi_{l',j'}}|0\rangle_{\textbf{D}_{l',j'}}+\frac{\psi_{l'+1,2j'+1}}{\psi_{l',j'}}|1\rangle_{\textbf{D}_{l',j'}}\right)
\end{align}
as the quantum state of the subsystem ${\mathscr{D}_{l,j}}$ for Eq.~\eqref{eq:mscrd}. We will then iteratively perform the transformation 
\begin{align}\label{eq:trans3}
|\psi_{l}\rangle\rightarrow|\psi_{l+1}\rangle,
\end{align}
where 
\begin{align}\label{eq:psi3}
|\psi_{l}\rangle=\sum_{j=0}^{2^l-1}\psi_{l,j}|\theta\rangle_{\mathscr{D}_{l,j}}\otimes|\mathscr{C}_{l,j}\rangle_{\mathscr{V}-\mathscr{D}_{l,j}}.
\end{align}
Here, we have defined $\mathscr{C}_{l,j}\equiv\mathscr{M}_{l,j}\cup\mathscr{B}_{l,j}$, with  $\mathscr{M}_{l,j}\equiv\{\textbf{M}_{0,0}\}\cup\left\{\textbf{M}_{l',j_{1:l'}}\big|1\leqslant l'\leqslant l-1\right\},$ and
\begin{subequations}\label{eq:BBb}
\begin{align}
\mathscr{B}'_{l,j}&=\left\{ \textbf{D}_{l',j'}\in\mathscr{A}_{l,\downarrow,j}\big|{j'}_{l'}=1\right\}\\
\mathscr{B}_{l,j}&=\mathscr{B}'_{l,j}\cup\{\textbf{U}_{l,j}\}.
\end{align}
\end{subequations}
$\mathscr{M}_{l,j}$ includes all ancestors of $\textbf{M}_{l,j}$ in the middle layers, and Eq.\eqref{eq:BBb} is the same as Eq.~\eqref{eq:BB}.

Note that $|\psi_0\rangle=|\theta\rangle_{\mathscr{D}}\otimes|\{\textbf{D}_{0,0}\}\rangle$. In Eq.~\eqref{eq:psi3}, quantum state of qubit set $\mathscr{D}_{l,j}$ and $\mathscr{V}-\mathscr{D}_{l,j}$ (all qubits not in $\mathscr{D}_{l,j}$) are expressed in the form of computational basis representation and activation representation, respectively.

By implementing parallel CNOT gates (see also Fig.~\ref{fig:skt3in} (b)-(c) and (e)-(f)), we have $\textbf{PCNOT}_l|\mathscr{C}_{l,j}\rangle_{\mathscr{V}-\mathscr{D}_{l,j}}=|\mathscr{C}_{l,j}\cup\{\textbf{M}_{l,j}\}\rangle_{\mathscr{V}-\mathscr{D}_{l,j}}$, and hence 

\begin{align}\label{eq:psi3p}
\textbf{PCNOT}_l|\psi_{l}\rangle=\sum_{j=0}^{2^l-1}\psi_{l,j}|\theta\rangle_{\mathscr{D}_{l,j}}\otimes|\mathscr{C}_{l,j}\cup\{\textbf{M}_{l,j}\}\rangle_{\mathscr{V}-\mathscr{D}_{l,j}}.
\end{align}
Then, if we apply $\textbf{CRT}_{l,j}$ on Eq.~\eqref{eq:psi3p}, only basis with label $j$ will be changed. This is because the routing is controlled on $\textbf{M}_{l,j}$. The basis with label $j$ can be rewritten as (see also Fig.~\ref{fig:skt3in} (c)-(d) and (f)-(h))

\begin{strip}
\rule{0.5\textwidth}{.4pt}
\begin{align}
&\psi_{l,j}\left(|\theta\rangle_{\mathscr{D}_{l,j}}\otimes|\mathscr{C}_{l,j}\cup\{\textbf{M}_{l,j}\}\rangle_{\mathscr{V}-\mathscr{D}_{l,j}}\right)
\notag\\
=&\psi_{l,j}|\theta\rangle_{\mathscr{D}_{l,j}-\{\textbf{D}_{l,j}\}}\otimes \left(\psi_{l+1,2j}/\psi_{l,j}|\mathscr{C}_{l,j}\cup\{\textbf{M}_{l,j}\}\rangle_{\mathscr{V}-\mathscr{D}_{l,j}-\{\textbf{D}_{l,j}\}}+\psi_{l+1,2j+1}/\psi_{l,j}|\mathscr{C}_{l,j}\cup\{\textbf{M}_{l,j},\textbf{D}_{l,j}\}\rangle_{\mathscr{V}-\mathscr{D}_{l,j}-\{\textbf{D}_{l,j}\}}\right)
\notag\\
=&\psi_{l+1,2j}|\theta\rangle_{\mathscr{D}_{l+1,2j}}\otimes|\mathscr{C}_{l,j}\cup\{\textbf{M}_{l,j}\}\rangle_{\mathscr{V}-\mathscr{D}_{l+1,2j}}+\psi_{l+1,2j+1}|\varphi\rangle_{\mathscr{D}_{l+1,2j+1}}\otimes|\mathscr{C}_{l,j}\cup\{\textbf{M}_{l,j},\textbf{D}_{l,j}\}\rangle_{\mathscr{V}-\mathscr{D}_{l+1,2j+1}}.
\end{align}
In set $\mathscr{C}_{l,j}$, $\textbf{U}_{l,j}$ is activated, and operation $\textbf{CRT}_{l,j}$ moves this activation to either $\textbf{U}_{l+1,2j}$ or $\textbf{U}_{l+1,2j+1}$, depending on whether $\textbf{D}_{l,j}$ is activated or not. Thus, it can be verified that
\begin{align}
&\textbf{CRT}_{l,j}|\mathscr{C}_{l,j}\cup\{\textbf{M}_{l,j}\}\rangle_{\mathscr{V}-\mathscr{D}_{l+1,2j}}=|\mathscr{C}_{l+1,2j}\rangle_{\mathscr{V}-\mathscr{D}_{l+1,2j}}\\
&\textbf{CRT}_{l,j}|\mathscr{C}_{l,j}\cup\{\textbf{M}_{l,j},\textbf{D}_{l,j}\}\rangle_{\mathscr{V}-\mathscr{D}_{l+1,2j+1}}=|\mathscr{C}_{l+1,2j}\rangle_{\mathscr{V}-\mathscr{D}_{l+1,2j+1}}.
\end{align}

\hspace{0.5\textwidth}\rule{0.5\textwidth}{.4pt}
\end{strip}

\begin{strip}
\rule{0.5\textwidth}{.4pt}

See also Fig.~\ref{fig:skt3in} (d)-(e) and (g)-(h)) for illustration. Accordingly, we have

\begin{align}
\textbf{PCRT}_l\textbf{PCNOT}_l|\psi_{l}\rangle=&\sum_{j=0}^{2^l-1}\psi_{l+1,2j}|\theta\rangle_{\mathscr{D}_{l+1,2j}}\otimes|\mathscr{C}_{l+1,2j}\rangle_{\mathscr{V}-\mathscr{D}_{l+1,2j}}+\psi_{l+1,2j+1}|\theta\rangle_{\mathscr{D}_{l+1,2j+1}}\otimes|\mathscr{C}_{l+1,2j+1}\rangle_{\mathscr{V}-\mathscr{D}_{l+1,2j+1}}
\notag\\
=&\sum_{j=0}^{2^{l+1}-1}\psi_{l,j}|\theta\rangle_{\mathscr{D}_{l,j}}\otimes|\mathscr{C}_{l,j}\rangle_{\mathscr{V}-\mathscr{D}_{l,j}}\notag\\
=&|\psi_{l+1}\rangle.
\end{align}
\hspace{0.5\textwidth}\rule{0.5\textwidth}{.4pt}
\end{strip}

Applying $\textbf{PCRT}_l\textbf{PCNOT}_l$ iteratively from $l=0$ to $l=n-1$, we obtain
\begin{align}
|\psi_n\rangle=\sum_{j=0}^{N-1}\psi_{j}|\theta\rangle_{\mathscr{D}_{n,j}}\otimes|\mathscr{C}_{n,j}\rangle_{\mathscr{V}-\mathscr{D}_{n,j}}.
\end{align}
In the last step, we perform $\textbf{DECODE}$, i.e. applying $r^\dag_{l,j}$ on $\textbf{D}_{l,j}$ conditioned on $\textbf{M}_{l,j}$ not activated. For basis with label $j$, at the middle sublayers, only qubits $\textbf{M}_{n-1,j}$, $\textbf{M}_{n-2,j_{1:n-1}}$, $\cdots$, $\textbf{M}_{0,j_1}$ are activated. These qubits are not in $\mathscr{D}_{n,j}$, so $|\theta\rangle_{\mathscr{D}_{l,j}}$ are uncomputed, and the final state is (see also Fig.~\ref{fig:skt3in} (h)-(i))
\begin{align}\label{eq:in3}
|\psi\rangle=\textbf{DECODE}|\psi_n\rangle=\sum_{j=0}^{N-1}\psi_{j}|\mathscr{C}_{n,j}\rangle_{\mathscr{V}}.
\end{align}
Eq.~\eqref{eq:in3} is similar to the one for the 2-qubit-per-node protocol. The only difference is that for basis $j$, all ancestors of $\textbf{U}_{n,j}$ in sublayers $(l,\bullet)$ are activated. In the next section, with a mild modification of the fanout phase, we can uncompute the QRAM while obtaining the target state in the output register.

\subsection{Fanout phase}
\begin{figure*}[h]
          \includegraphics[width=2\columnwidth]{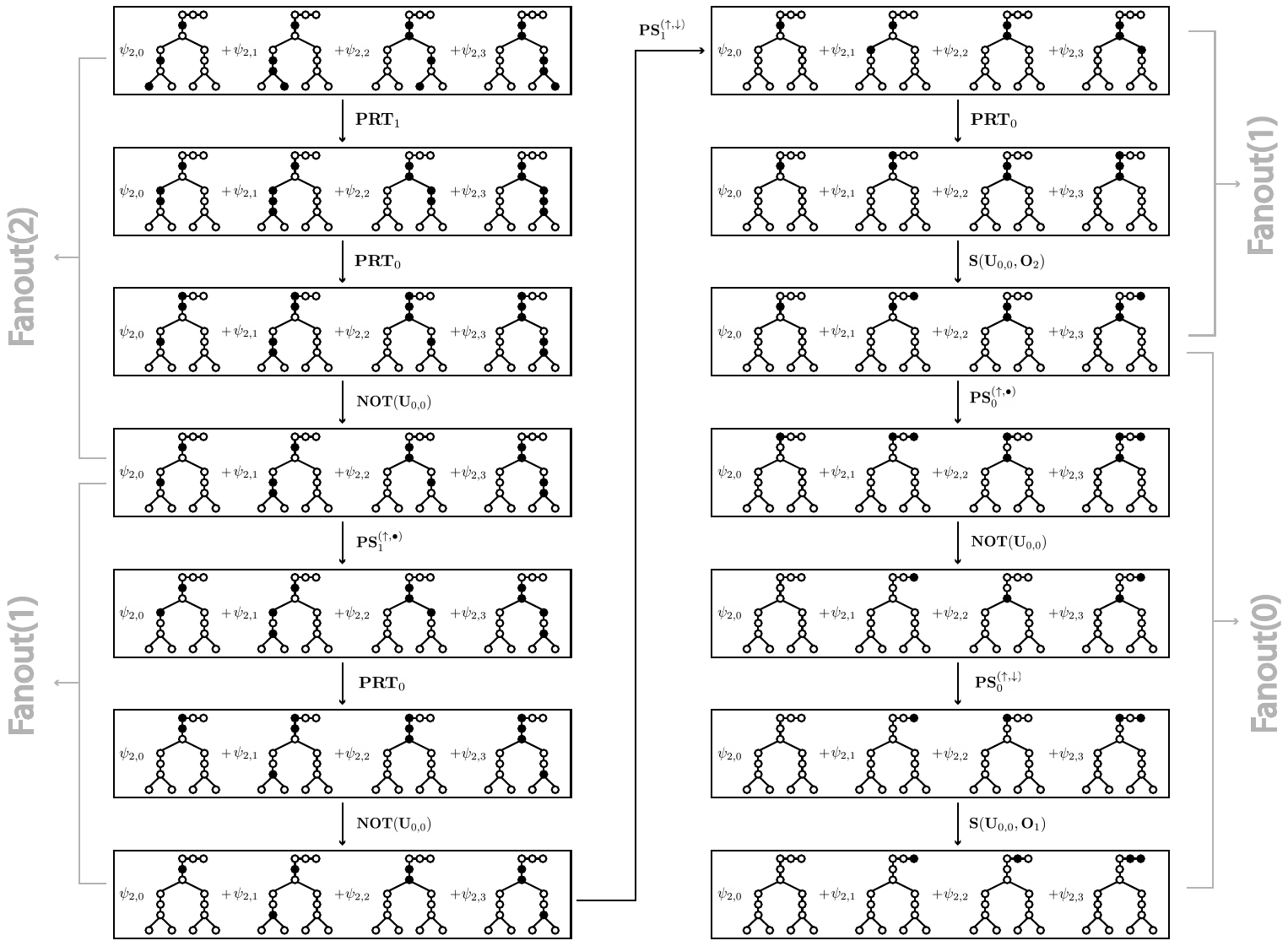}
       \caption{Sketch of the fanout process of 3-qubit-per-node protocol for $n=2$ case. Hollow and solid circles represent qubits at quantum state $|0\rangle$ and $|1\rangle$ respectively.} \label{fig:skt3out}
\end{figure*}

We now discuss the fanout phase of our algorithms, which corresponds to lines 6-8 in Algorithm.~\ref{alg:sp3}. An example for $n=2$ is also illustrated in Fig.~\ref{fig:skt3out}. 
Let 
\begin{align}
\mathscr{C}'_{l,j}=\mathscr{M}_{l,j}\cup\mathscr{B}'_{l,j},
\end{align}
with $\mathscr{B}'_{l,j}$ defined in Eq.~\eqref{eq:BB}, it can be verified that 
\begin{align}\label{eq:fan3p}
\textbf{NOT}(\textbf{U}_{0,0})\textbf{PCRT}_{1}\textbf{PCRT}_{2}\cdots\textbf{PCRT}_{n-1} |\mathscr{C}_{n,j}\rangle=|\mathscr{C}'_{n,j}\rangle.
\end{align}
In other words, performing parallel controlled routing from $l=n-1$ to $l=0$ transfers the excitation at layer $(n,\uparrow)$ to $\textbf{U}_{0,0}$, which can be uncomputed by an extra not gate. Our strategy is to perform the following transformation 
\begin{align}\label{eq:fan3tr}
&|\mathscr{C}_{l,j_{1:l}}'\rangle\otimes|0\cdots 0 j_{l+1}\cdots j_n\rangle_{\text{out}}\notag\\
\longrightarrow&|\mathscr{C}_{l,j_{1:l-1}}'\rangle\otimes|0\cdots 0 j_{l}\cdots j_n\rangle_{\text{out}}.
\end{align}
iteratively. For basis $|\mathscr{C}_{l,j_{1:l}}'\rangle$, we can also deterministically route the activation at layer $(l,\bullet)$ to $\textbf{U}_{0,0}$, and uncompute it with a NOT gate, i.e.
\begin{align}\label{eq:fan30}
&\textbf{NOT}(\textbf{U}_{0,0})\textbf{PCRT}_{1}\textbf{PCRT}_{2}\cdots\textbf{PCRT}_{l-1}\textbf{PS}_{l}^{(\uparrow,\bullet)}|\mathscr{C}_{l,j_{1:l}}'\rangle\notag\\
=&|\mathscr{C}_{l,j_{1:l}}'-\{\textbf{M}_{l,j_{1:l}}\}\rangle.
\end{align}
Moreover, in analogy to the 2-qubit-per-node protocol, we can route the state $|j_l\rangle$ from layer $(l,\downarrow)$ to qubit $\textbf{O}_{l}$ in the output register by

 \begin{align}\label{eq:fan31}
&\textbf{S}(\textbf{U}_{0,0},\textbf{O}_{l})\textbf{PCRT}_{1}\textbf{PCRT}_{2}\cdots\textbf{PCRT}_{l-1}\textbf{PS}_{l}^{(\uparrow,\downarrow)}\notag\\
&\times|\mathscr{C}_{l,j_{1:l}}'-\{\textbf{M}_{l,j_{1:l}}\}\rangle\otimes|0\cdots0 j_{l+1}\cdots j_{n}\rangle_{\text{out}}\notag\\
=&|\mathscr{C}_{l,j_{1:l-1}}'\rangle\otimes|0\cdots0 j_{l}\cdots j_{n}\rangle_{\text{out}}.
\end{align}
We can start the operation in Eq.~\eqref{eq:fan31} after the operation in Eq.~\eqref{eq:fan30} has finished the $\textbf{PCRT}_{l-2}$, and two operations will not affect each other.  With an abuse of notation, we also define this process as $\textbf{Fanout}(l-1)$ (for $1\leqslant l\leqslant n$), 
which performs the transformation claimed in Eq.~\eqref{eq:fan3tr}. We also define $\textbf{Fanout}(n)$ as the process corresponding to Eq.~\eqref{eq:fan3p}. By implementing $\textbf{Fanout}(n), \textbf{Fanout}(n-1),\cdots, \textbf{Fanout}(0)$ iteratively, we can uncompute the QRAM, while preparing the target state at output register. Similar to the 2-site-per-node protocol, while implementing $\textbf{Fanout}(l)$ sequentially is time costly, we can start the next Fanout operation before the current operation is finished. More specifically, we can start $\textbf{Fanout}(l)$, idle for $5$ steps, and then start $\textbf{Fanout}(l-1)$. In this way, operations $\textbf{Fanout}(l)$ and $\textbf{Fanout}(l-1)$ will not affect each other, and the total runtime is $O(n)$.

\subsection{Clifford$+T$ decomposition}\label{sec:ct}
\subsubsection{Decomposition protocol and error analysis}
Among all elementary single- and two-qubit gates, only rotations $\textbf{R}_{l,j}$ and controlled rotations $\overline{\textbf{CR}}_{l,j}$ have decomposition errors, while all other elementary gates can be ideally constructed with constant number of Clifford and $T$ gates.

According to~\cite{Selinger.12}, given an arbitrary z-rotation $R_z(\alpha)$ and accuracy $\varepsilon>0$, we can always construct a single qubit rotation $R_z(\alpha,\varepsilon)$ with $O(\log(1/\varepsilon))$ depth of $H$ and $T$ gates, such that $\|R_z(\alpha,\varepsilon)-R_z(\alpha)\|\leqslant\varepsilon$. For y-rotation $R_y(\beta)$, the result is similar.
Moreover, we can always decompose each $r_{l,j}$ into the concatenation of a y-rotation and a $z$-rotation $r_{l,j}=R_z(\alpha_{l,j})R_y(\beta_{l,j})$. We approximate $\textbf{R}_{l,j}$ with the following quantum circuit

\begin{centering}
\quad \begin{quantikz}
 \textbf{D}_{l,j}&\gate{r_{l,j}(\varepsilon)}&\qw&
 \end{quantikz}  
$\equiv$\quad
%
 
  \begin{quantikz}
 \textbf{D}_{l,j}&\gate{R_y\left(\frac{\beta_{l,j}}{2},\frac{\varepsilon}{4}\right)}&\gate{R_x(\pi)}&\gate{R_y\left(\frac{\beta_{l,j}}{2},\frac{\varepsilon}{4}\right)^\dag}&\qw\cdots
 \end{quantikz} 
 
   \begin{quantikz}
\,\,\,\cdots&\gate{R_z\left(\frac{\alpha_{l,j}}{2},\frac{\varepsilon}{4}\right)^\dag}&\gate{R_x(\pi)}&\gate{R_z\left(\frac{\alpha_{l,j}}{2},\frac{\varepsilon}{4}\right)}&\qw
 \end{quantikz} 
 \end{centering}
 \vspace{.5cm}
 
\noindent which we denote as  $\textbf{R}_{l,j}(\varepsilon)$.
 
 Note that $R_z\left(\frac{\alpha_{l,j}}{2},\frac{\varepsilon}{4}\right)^\dag$ can be constructed by inverse the $H,T$ gate sequence of $R_z\left(\frac{\alpha_{l,j}}{2},\frac{\varepsilon}{4}\right)$, then replace $T$ and $H$ by $T^\dag$ and $H^\dag$, and similar for $y$-rotation. $R_x(\pi)$ takes $O(1)$ gate count, and $\|\textbf{R}_{l,j}-\textbf{R}_{l,j}(\varepsilon)\|\leqslant\varepsilon$. The reason for using this decomposition is that together with the controlled rotation introduced below, qubits in $\mathscr{D}_{l,j}$ can be fully uncomputed after implementing $\overline{\textbf{CR}}_{l,j}$. To be specific, $\overline{\textbf{CR}}_{l,j}$ is approximated by 

\vspace{.5cm}
  \begin{centering}
 \begin{quantikz}
 \textbf{M}_{l,j}&\octrl{1}&\qw\\
 \textbf{D}_{l,j}&\gate{r_{l,j}(\varepsilon)^\dag}&\qw&
 \end{quantikz}  $\equiv$

 \begin{quantikz}
 \textbf{M}_{l,j}&\qw&\octrl{1}&\qw &\qw\cdots\\
 \textbf{D}_{l,j}&\gate{R_z(\frac{\alpha_{l,j}}{2},\frac{\varepsilon}{4})^\dag}&\gate{R_x(\pi)}&\gate{R_z(\frac{\alpha_{l,j}}{2},\frac{\varepsilon}{4})}&\qw\cdots
 \end{quantikz}  
 
 \begin{quantikz}
 \,\,\,\,\,\,\cdots &\qw&\octrl{1}&\qw&\qw&\qw\\
 \,\,\,\,\,\,\cdots &\gate{R_y(\frac{\beta_{l,j}}{2},\frac{\varepsilon}{4})}&\gate{R_x(\pi)}&\gate{R_y(\frac{\beta_{l,j}}{2},\frac{\varepsilon}{4})^\dag}&\qw&\qw
 \end{quantikz}  
 \end{centering}
 \vspace{.5cm}

\noindent which we denote as  $\overline{\textbf{CR}}_{l,j}(\varepsilon)$.

In our Clifford$+T$ circuit implementation, we just perform the following replacement in the fanin phase
\begin{align}
\textbf{R}_{l,j}\rightarrow\textbf{R}_{l,j}(\varepsilon_l), \quad \overline{\textbf{CR}}_{l,j}\rightarrow\overline{\textbf{CR}}_{l,j}(\varepsilon_l).
\end{align}
To analyze the decomposition accuracy, we define 
\begin{eqnarray}\notag
U_l= \left\{
\begin{array}{lll}
r_{0,0}\otimes \mathbb{I}_{n-1},\quad     &    & l=0 \\
\sum_{j=0}^{2^{l}-1}|j\rangle\langle j|\otimes r_{l,j}\otimes \mathbb{I}_{n-l-1},   &  & 1\leqslant l\leqslant n-1\label{eq:parent}\\  
\end{array} \right. \ 
\end{eqnarray}
and 
\begin{eqnarray}\notag
U_l(\varepsilon_l)= \left\{
\begin{array}{lll}
r_{0,0}(\varepsilon_0)\otimes \mathbb{I}_{n-1},\quad     & l=0 \\
\sum_{j=0}^{2^{l}-1}|j\rangle\langle j|\otimes r_{l,j}(\varepsilon_l)\otimes \mathbb{I}_{n-l-1},    & 1\leqslant l\leqslant n-1\label{eq:parent}\\  
\end{array} \right. \ 
\end{eqnarray}
where $\mathbb{I}_{m}$ is the $m$-qubit identity matrix. It can be verified that for ideal and Clifford$+T$ implementations, the final state of the output register is equivalent to
\begin{subequations}\label{eq:uuct}
\begin{align}
|\psi\rangle&=U_{n-1}\cdots U_1U_0|0\rangle^{\otimes n}\\
|\psi^{(\text{CT})}\rangle&=U_{n-1}(\varepsilon_{n-1})\cdots U_1(\varepsilon_1)U_0(\varepsilon_0)|0\rangle^{\otimes n}
\end{align}
\end{subequations}
respectively, while the QRAM has been uncomputed for both cases. We note that Eq.~\eqref{eq:uuct} is only an expression of the final state, and does not represent the actual implementation process. Because $\|U_l-U_l(\varepsilon_l)\|\leqslant\varepsilon_l$, according to the triangular inequality, we have 

\begin{align}\label{eq:cte}
\left\||\psi\rangle-|\psi^{(\text{CT})}\rangle\right\|\leqslant\sum_{l=0}^{n-1}\|U_l-U_l(\varepsilon_l)\|\leqslant\sum_{l=0}^{n-1}\varepsilon_{l}.
\end{align}
Based on Eq.~\eqref{eq:cte}, to achieve a given accuracy $\left\||\psi\rangle-|\psi^{(\text{CT})}\right\|\leqslant\varepsilon$, it suffices to set 
\begin{align}
\varepsilon_l=\varepsilon/2^{n-l}.
\end{align}

\subsubsection{Circuit complexity}
Below, we analyze the Clifford$+T$ circuit complexity based on the decomposition protocol above.

$\\$
\noindent\textbf{Clifford$+T$ gate count.}  Each rotation $\textbf{R}_{l,j}(\varepsilon_l)$ or controlled-rotation $\textbf{CR}_{l,j}(\varepsilon_l)$ accounts for $O(\log1/\varepsilon_l)=O(\log (2^{n-l}/\varepsilon))$ gate count. So $\textbf{PR}$ and $\textbf{P}\overline{\textbf{CR}}$ accounts for totally $\sum_{l=1}^{n} 2^l\times O(\log(2^{n-l}/\varepsilon))=O(N\log(1/\varepsilon))$ gate count. Other operations during the implementation can be realized without decomposition errors, and account for $O(N)$ gate count. Therefore, the total Clifford$+T$ gate count is $O(N\log(1/\varepsilon))$.

$\\$
\noindent\textbf{Clifford$+T$ depth.} The decomposed parallel rotation and parallel controlled-rotation accounts for $\max_{l}O(\log(2^{n-l}/\varepsilon))=O(\log(2^{n}/\varepsilon))=O(n+\log(1/\varepsilon))$ circuit depth. Other operations during the implementation account for total $O(n)$ depth. So the total Clifford$+T$ depth is $O(n+\log(1/\varepsilon))$. 

$\\$
\noindent\textbf{Clifford$+T$ STA.} The STA is more involved. We first consider the naive implementation of Algorithm.~\ref{alg:sp3}. During \textbf{ENCODE}, all single-qubit rotations $\textbf{R}_{l,j}$ are implemented simultaneously and remain activated at least until the finish of \textbf{DECODE}. In this process, each qubit is activated for time $O(n+\log(1/\varepsilon))$. This leads to a total  STA $O(N(n+\log(1/\varepsilon)))$. 

 \begin{figure*}[!t]
 \centering
          \includegraphics[width=1.8\columnwidth]{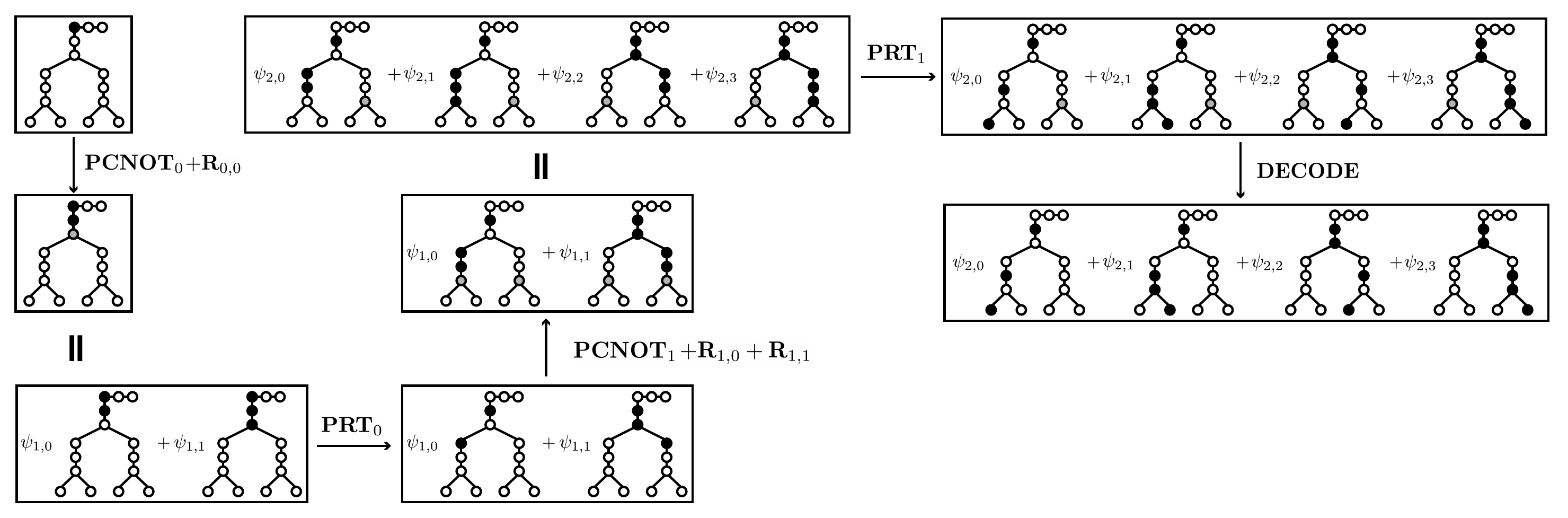}
       \caption{Sketch of the fanin process of 3-qubit-per-node protocol for $n=2$ case, with STA optimized. Hollow and solid circles represent qubits at quantum state $|0\rangle$ and $|1\rangle$ respectively.}\label{fig:skt3insta}
\end{figure*} 

Fortunately, instead of implementing rotations within \textbf{ENCODE} simultaneously, we can delay the implementation most $\textbf{R}_{l,j}$. Each $\textbf{R}_{l,j}$ can be implemented \textit{as late as possible}, such that when $\textbf{R}_{l,j}$ finish, the subsequent routing operation $\textbf{RT}_{l,j}$ begin. In this way, the STA of the process is significantly reduced. In Fig.\ref{fig:skt3insta}, 
 we also sketch the fanin process when STA is optimized.

We now evaluate the total STA of the optimized scheme. Let us first consider qubit $\textbf{D}_{l,j}$ at the lower sublayer. It is activated for time $O(\log(2^{n-l}/\varepsilon))=O((n-l)\log(1/\varepsilon))$  during encoding,  time $O(n-l)$ during the routing process of fanin, time $O(\log(2^{n-l}/\varepsilon))=O((n-l)\log(1/\varepsilon))$ for decoding, and $O((n-l)\log(1/\varepsilon))$ for fanout. So it is activated for total time  $O((n-l)\log(1/\varepsilon))$. For other qubit $\textbf{U}_{l,j}$ and $\textbf{M}_{l,j}$, they are all activated for total time $O(n-l)$. Summing the activated time for all qubits, the total STA is 
\begin{align}
\text{STA}&=\sum_{l=1}^{n}2^l\times O((n-l)\log(1/\varepsilon))+\sum_{l=1}^n2^{l}\times O(n-l)\notag\\
&=O(N\log(1/\varepsilon)).
\end{align}
The circuit complexity is summarized in Table.~\ref{tab:1}. 

\section{Application to Block-encoding}\label{sec:be}
Block-encoding enables the embedding of a general matrix $M$ into a unitary with higher dimension. It is a basic operation in quantum algorithm, and together with quantum singular-value transformation, they can unify most of the fault-tolerant quantum algorithms~\cite{Martyn.21}. Specifically, we say a unitary $W$ is the $(\alpha,n_{\text{anc}},\varepsilon)$-block-encoding of $M$ if $\|\langle0^{\text{anc}}|W|0^{\text{anc}}\rangle-M/\alpha\|\leqslant\varepsilon$, for some normalization factor $\alpha$ and ancillary qubit number $n_{\text{anc}}$.  Below, we show that techniques introduce in previous sections are also applicable for the robust realization of block-encoding.

\subsection{block-encoding general matrix}\label{sec:beg}
We begin with the block-encoding of a general unstructured matrix
$M=\sum_{j,k=0}^{N-1}M_{j,k}|j\rangle\langle k|$. Following the protocol in~\cite{Clader.22}, we introduce two subsystems, each with $n$ qubits. 
We define $|M\rangle=\sum_{j=0}^{N-1}\frac{\|M_{j,\cdot}\|_F}{\|M\|_F}|j\rangle$ and $|M_j\rangle=\sum_{k=0}^{N-1}\frac{M_{j,k}^*}{\|M_j,\cdot\|_F}|k\rangle$, where $\|\cdot\|_F$ is the Frobenius norm. We introduce three unitaries, $\text{SWAP}, U_R,$ and $U_L$ satisfying the following
\begin{align}
\text{SWAP}|j\rangle|k\rangle&=|k\rangle|j\rangle,\label{eq:bs}\\
U_{L}|k\rangle|0^{n}\rangle&=|k\rangle|M\rangle,\label{eq:bul}\\
U_R|j\rangle|0^{n}\rangle&=|j\rangle|M_j\rangle.\label{eq:bur}
\end{align}
It can then be verified that $W\equiv U_R^{\dag}\text{SWAP}U_{L}$ is the block-encoding of $M$ with normalization factor $\|M\|_F$. Eq.~\eqref{eq:bs} can be realized in constant layer of elementary gates with the following circuit.
\begin{center}
\begin{quantikz}
& \swap{4} &\qw &  \qw & \qw &\qw  \\
& \qw &\swap{4} &  \qw & \qw &\qw  \\
& \qw &\qw &  \swap{4} & \qw &\qw  \\
\wave&&&&&&\\
& \swap{} & \qw & \qw & \qw & \qw \\
& \qw & \swap{} &\qw &\qw & \qw\\
& \qw & \qw &\swap{} &\qw & \qw
\end{quantikz}
\end{center}
\vspace{.2cm}
$U_{L}$ is just a state preparation unitary, and we may assume that it is realized with our 3-qubit-per-node protocol.  $U_R$ is a controlled state preparation, which can be considered as the generalization of the QRAM operation. A general multi-qubit-controlled-unitaries, can be realized with bucket-brigade approach~\footnote{Although~\cite{Zhang.22} uses 2-qubit-per-node protocol, the revision to 3-qubit-per-node approach is straightforward.}, together with a layer of (totally $N$ number of) single-qubit controlled unitaries (Algorithm 4,5 in~\cite{Zhang.22}, see also Lemma 7 in~\cite{Zhang.24}). Each controlled-state-preparation can be realized by the approach in Sec.~\ref{sec:3bit}, and has the infidelity scaling $O(\varepsilon n^2)$. Due to the same noise-robustness mechanism in~\cite{Hann.21} and this work, errors will not propagate between different branches of controlled state preparation in most cases, and hence the total infidelity scaling of implementing $U_R$ remains  to be $O(\varepsilon n^2)$. Moreover, based on Lemma 7 and relevant discussions in~\cite{Zhang.24}, the circuit depth, gate count and STA for $U_{R}$ are $O(n+\log(1/\varepsilon))$, $O(N^2\log(1/\varepsilon))$, and $O(N^2\log(1/\varepsilon))$ respectively, given totally $O(N^2)$ number of ancillary qubits. The same infidelity scaling and gate complexity is also applied for implementing $W$, i.e. the $(\|M\|_F,n_{\text{anc}},\varepsilon)$-block-encoding of $M$, for some $n_{\text{anc}}=O(N^2)$.

\subsection{Block-encoding LCU}
 LCU~\cite{Long.06,Long.11,Childs.12} is a much less costly model compared to general matrices, yet has broad applications. We consider the following matrix form

\begin{align}\label{eq:belcu}
H=\sum_{p=1}^P\alpha_{p}u_p,
\end{align}
where $\alpha_p>0$ and $u_p$ are $O(1)$-local unitaries, i.e. applied at a constant number of qubits. Eq.~\eqref{eq:belcu} can represent most of the quantum many-body systems with local interactions. The block-encoding of $H$ can be realized by the operation 
\begin{align}
W\equiv(\text{SP}^\dag\otimes \mathbb{I})\text{SELECT}(\text{SP}\otimes \mathbb{I}).
\end{align}
Here, SP is a state preparation unitary applying at the ancillary system, which satisfies SP$|0^{\text{anc}}\rangle=\sum_{p=1}^P\sqrt{\alpha_p/\alpha}|p\rangle$, where $\alpha=\sum_{p=1}^P\alpha_p$. SELECT$=\sum_{p=1}^P|p\rangle\langle p|\otimes u_p$ is the select operator, and similar to the discussion in previous section, this multi-qubit-controlled-unitary can be realized by a bucket-brigade approach with one layer of controlled-$u_p$ gates. Note that due to the locality assumption, each controlled-$u_p$ can be realized by a constant layer of single- and two-qubit gates. So the infidelity scaling, circuit depth, gate count and STA of the SELECT operation are $O(\varepsilon\log^2(P))$, $O(\log(P)+\log(1/\varepsilon))$, $O(P\log(1/\varepsilon))$ and $O(P\log(1/\varepsilon))$ respectively. Combining the implementation of state preparation, the $(\alpha,n_{\text{anc}},\varepsilon)$-block-encoding of $H$ also has the same performance to SELECT, with $n_{\text{anc}}=O(P)$. 

The improvement of noise robustness can significantly reduce the resources required for early-fault-tolerant quantum computing. We take the Hamiltonian simulation of a geometrically local Hamiltonian (e.g. Ising model, Heisenberg models) as an example. For an $n$-qubit system, we have $P=O(n)$. According to the discussions above, if we expect the total accuracy to achieve $\varepsilon$ under noise, the infidelity of each elementary gate is required to be $O(\varepsilon/\log^2(n))$, as opposed to $O(\varepsilon/n)$ for conventional methods. Accordingly, when performing error-correction~\cite{Fowler.12}, the code distance for each logic qubit can be exponentially reduced from $O(\text{polylog}(n))$ to $O(\text{polyloglog}(n))$, compared to other depth-optimal or few-ancillary methods~\cite{Long.01,Grover.02,Mottonen.05,Low.24,Zhang.21,Clader.22,Sun.21,Rosenthal.21,Zhang.22,Yuan.22,Gui.23,Zhang.24}. This level of improvement is applied for both dynamical simulation~\cite{Low.19} (assuming evolution time is independent of $n$) and ground energy estimation~\cite{Lin.20} (assuming accuracy is independent of $n$). 

\vspace{.2cm}
\section{Conclusion and Discussion}\label{sec:cd}
We have proposed practical, robust, and optimal approaches to quantum state preparation. The technique is also applicable to the block-encoding general matrices and LCU.  The approaches have infidelities scale polylogarithmically with data size, and at the same time achieve state-of-the-art circuit complexities. So it is particularly useful for near-term and early fault-tolerant quantum devices. 

While we have only considered the Pauli depolarization channel here,  it is expected that the protocol is  robust for general quantum noise models (e.g. dephasing, decaying), in case they are not catastrophic errors applied globally. Moreover, the robustness mechanism here is applicable to other type of data-loading process, such as 
  sparse quantum state preparation~\cite{Zhang.22,Mao.24,Luo.24} sparse-access input model~\cite{Berry.07} and function loading~\cite{Gabriel.21,Rattew.22}. 
  In the experimental aspect, our protocol is directly implementable in state-of-the-art quantum platforms, and serves as a promising candidate for future quantum data center~\cite{Liu.22,Liu.23}.
 

 \vspace{.2cm}
\noindent \textbf{Acknowledgement} The author thanks Alexander Denzel, Connor T. Hann and Xiao Yuan for helpful discussions. This work is supported by National Natural Science Foundation of China (No. 12405013), and Open Fund of Key Laboratory of Atomic and Subatomic Structure and Quantum Control (Ministry of Education).

\bibliographystyle{IEEEtran}


\onecolumn

\begin{appendices}

\section{Proof of the robustness for 2-qubit-per-node protocol}\label{sec:nr}
\subsection{noise model}
As explained in the main text, state preparation protocol contains totally $O(n)$ layers of quantum circuit.
We can abstractly expresse the quantum circuit as $\prod_{m=1}^MU_{m}|\psi_{\text{ini}}\rangle=|\psi\rangle$, where $U_{m}$ is the $m$th layer of single- and two-qubit gates. The specific form of $U_{m}$ depends on how we decompose the operations (e.g. elementary routing and control rotation operations), but we typically have $M=O(n)$. In practice, we should deal with mixed state due to the existence of noise, so we also define the corresponding unitary channels as $\mathcal{U}_{m}[\cdot]=U_{m}[\cdot] U_{m}^\dag$. Let 
\begin{align}
\mathcal{U}=\mathcal{U}_{M}\circ\cdots\circ\mathcal{U}_{2}\circ\mathcal{U}_{1}
\end{align}
be the ideal evolution,
 we have $\mathcal{U}[\rho_{\text{ini}}]=\rho_{\text{end}}$, where $\rho_{\text{ini}}=|\psi_{\text{ini}}\rangle\langle\psi_{\text{ini}}|$, and $\rho_{\text{end}}=|\psi_{\text{end}}\rangle\langle\psi_{\text{end}}|$ are initial and ideal output state. Let $\rho_{\text{id}}=|\psi\rangle\langle\psi|$ be the target state of the quantum state preparation, we have  $\rho_{\text{id}}=\text{Tr}_{\text{qram}}[\rho_{\text{end}}]$, where $\text{Tr}_{\text{qram}}$ is the partial trace over the QRAM.

We then introduce the local depolarization noise model. We define
\begin{align}\label{eq:eq}
\mathcal{E}_q=(1-\varepsilon)\mathcal{I}+\frac{1}{3}\varepsilon \left(\mathcal{X}_q+ \mathcal{Y}_q+ \mathcal{Z}_q\right)
\end{align}
as the noisy quantum channel applied at qubit $q$, where $\varepsilon\in(0,1)$ is the error probability, $\mathcal{I}[\rho]=\rho$, $\mathcal{X}[\rho]=X_q\rho X_q$, $\mathcal{Y}[\rho]=Y_q\rho Y_q$, $\mathcal{Z}[\rho]=Z_q\rho Z_q$, are Pauli $I$, $X$, $Y$ and $Z$ channels applied at qubit $q$ respectively.  After the implementation of each layer of quantum circuit $\mathcal{U}_{m}$, $\mathcal{E}_q$ is applied at all qubits in the system. In other words, let $\mathcal{E}\equiv\prod_{q\in \mathscr{V}}\mathcal{E}_q$, where $\mathscr{V}$ is the set of all qubits in both QRAM and output register, the ideal channel $\mathcal{U}_{m}$ is replaced by the noisy channel $\tilde{\mathcal{U}}_{m}=\mathcal{E}\circ\mathcal{U}_{m}$. So the noisy quantum state preparation can be described by the following quantum channel 
\begin{align}\label{eq:errU}
\tilde{\mathcal{U}}\equiv\tilde{\mathcal{U}}_{M}\circ\cdots\circ\tilde{\mathcal{U}}_{2}\circ\tilde{\mathcal{U}}_{1}.
\end{align}

\subsection{Linear combination of unitary evolutions}
We then show how to decompose $\tilde{\mathcal{U}}$ into the linear combination of unitary evolutions.
We first rewrite $\mathcal{E}$ as the linear combination of all possible qubit distribution of error
\begin{align}
\mathcal{E}\equiv\sum_{Q\in \text{Power}(\mathscr{V})}\mathcal{E}_{Q}\equiv\sum_{Q\in \text{Power}(\mathscr{V})}p_Q\mathcal{D}_Q,
\end{align}
 where $\text{Power}(\mathscr{V})$ is the power of $\mathscr{V}$, i.e. all possible subset of all qubits.
Moreover, $\mathcal{D}_{q}=\mathcal{X}_q+\mathcal{Y}_q+\mathcal{Z}_q$ represents the depolarization part of Eq.~\eqref{eq:eq}, and
$\mathcal{D}_Q=\prod_{q\in Q}\mathcal{D}_q$, $p_Q=(1-\varepsilon)^{|\mathscr{V}|-|Q|}\varepsilon^{|Q|}$. Here, $\mathcal{D}_Q$ represents that errors are applied at qubits in set $Q$ while all qubits not in $Q$ is free of errors. The probability distribution $p_{Q}$ is normalized, and decreases with $|Q|$. 

Then, let $\boldsymbol{Q}\equiv[Q_1,\cdots,Q_M]$ be a vector of qubit set for some $Q_m\in \text{Power}(\mathscr{V})$. $\boldsymbol{Q}$ describes a specific space-time configuration of the depolarization error. More specifically, we define
\begin{align}\label{eq:Uq0}
\tilde{\mathcal{U}}(\boldsymbol{Q})&=\mathcal{D}_{Q_M}\circ\mathcal{U}_{M}\circ\cdots\circ\mathcal{D}_{Q_2}\circ\mathcal{U}_{2}\circ\mathcal{D}_{Q_1}\circ\mathcal{U}_{1},
\end{align}
and $p_{\boldsymbol{Q}}=\prod_{m=1}^Mp_{Q_{m}}$. Let $\mathscr{Q}=\left\{[Q_1,\cdots,Q_M]\big| Q_m\in \text{Power}(\mathscr{V})  \text{ for all } 1\leqslant m\leqslant M\right\}$ be all possible space-time configurations, we can rewrite $\tilde{\mathcal{U}}$ in Eq.~\eqref{eq:errU} as
\begin{align}\label{eq:uqq}
\tilde{\mathcal{U}}=\sum_{\boldsymbol{Q}\in\mathscr{Q}}p_{\boldsymbol{Q}}\tilde{\mathcal{U}}(\boldsymbol{Q}).
\end{align}
We further decompose each $\tilde{\mathcal{U}}(\boldsymbol{Q})$ into the linear combination of unitary evolutions. Recall that in Eq.~\eqref{eq:Uq0}, each depolarization $\mathcal{D}_{Q_m}$ is the linear combination of three unitary channels. Let $\mathscr{P}_{Q_m}=\left\{\prod_{q\in Q_m}\mathcal{P}_q \Large| \mathcal{P}_q\in\{\mathcal{X}_q,\mathcal{Y}_q,\mathcal{Z}_q\}\right\}$, we have
\begin{align}
\mathcal{D}_{Q_m}=\frac{1}{\left|\mathscr{P}_{Q_m}\right|}\sum_{\mathcal{P}\in\mathscr{P}_{Q_m}}\mathcal{P},
\end{align}
where $\left|\mathscr{P}_{Q_m}\right|=3^{|Q_m|}$.
Let $[\mathcal{P}_1,\mathcal{P}_2,\cdots,\mathcal{P}_M]$ be the polarization configuration of errors, we define all possible $[\mathcal{P}_1,\mathcal{P}_2,\cdots,\mathcal{P}_M]$ under a space-time configuration $\boldsymbol{Q}$ as $\mathscr{P}_{\boldsymbol{Q}}\equiv\left\{[\mathcal{P}_{1},\mathcal{P}_{2},\cdots,\mathcal{P}_{M}]\big| \mathcal{P}_{m}\in\mathscr{P}_{Q_m}\right\}$. $\tilde{\mathcal{U}}(\boldsymbol{Q})$ can therefore be decomposed as
\begin{align}
\tilde{\mathcal{U}}(\boldsymbol{Q})=\frac{1}{\left|\mathscr{P}_{\boldsymbol{Q}}\right|}\sum_{c\in\mathscr{P}_{\boldsymbol{Q}}}\tilde{\mathcal{U}}(c),
\end{align}
where $c$ represents a specific space-time-polarization configuration of error, and
\begin{align}\label{eq:u1}
\tilde{\mathcal{U}}([\mathcal{P}_{1},\mathcal{P}_{2},\cdots,\mathcal{P}_{M}])\equiv\mathcal{P}_{M}\mathcal{U}_M\cdots\mathcal{P}_{2}\mathcal{U}_2\mathcal{P}_{1}\mathcal{U}_1.
\end{align}
Because each $\mathcal{P}_m$ is a unitary channel,  Eq.~\eqref{eq:u1} is also a unitary channel. The total noisy evolution can then be decomposed as the linear combination of unitary evolutions as 
\begin{align}\label{eq:ulcue}
\tilde{\mathcal{U}} = \sum_{\boldsymbol{Q}\in\mathscr{Q}}\sum_{c\in\mathscr{P}_{\boldsymbol{Q}}}p_c\tilde{\mathcal{U}}(c)
\end{align}
for some $p_c = p_{\boldsymbol{Q}}/|\mathscr{P}_{\boldsymbol{Q}}|$. Let 
\begin{align}\label{eq:roc}
\tilde{\rho}_{\text{out}}(c)=\text{Tr}_{\text{qram}}\left[\tilde{\mathcal{U}}(c)[\rho_{\text{ini}}]\right],
\end{align}
the final noisy output state is therefore $\tilde{\rho}_{\text{out}}=\sum_c p_c\tilde{\rho}_{\text{out}}(c)$, where the sum is over all possible space-time-polarizaiton configurations. We denote  $\text{Fid}(A,B)$ as the fidelity between two density matrices $A$ and $B$, the total state preparation fidelity is just $F\equiv\text{Fid}(\rho_{\text{id}},\tilde{\rho}_{\text{out}})$. Due to the concavity of fidelity, we have
\begin{align}
F&\geqslant \sum_{\boldsymbol{Q}\in\mathscr{Q}}\sum_{c\in\mathscr{P}_{\boldsymbol{Q}}}p_c\text{Fid}[\rho_{\text{id}},\tilde{\rho}_{\text{out}}(c)]\notag\\
&=\mathbb{E}\left[\text{Fid}(\rho_{\text{id}},\tilde{\rho}_{\text{out}}(c))\right],\label{eq:cov}
\end{align}
where $\mathbb{E}[\cdot]$ represents the expectation value with $c$ sampled according to $p_c$. The remaining of this section is to study the unitary evolution under space-time-depolarization configuration $c$, and estimate Eq.~\eqref{eq:cov}.

\subsection{Definition of good branch}
Before discussing the infidelity of $\tilde{\rho}_{\text{out}}(c)$, we give the definition of \textit{good} branch and related terminologies that are useful. To begin with, we define the parent of each node as
\begin{eqnarray}
\text{Parent}\left[\boldsymbol{X}\right]= \left\{
\begin{array}{ccl}
\textbf{O}_{l+1}      &    & \boldsymbol{X}=\textbf{O}_{l} \text{ for } 1\leqslant l\leqslant n-1\\
\textbf{O}_{0}      &    & \boldsymbol{X}=\textbf{U}_{0,1} \\
\textbf{D}_{(l-1,\lceil j/2\rceil)}    &  & \boldsymbol{X}=\textbf{U}_{(l,j)} \text{ for some } l\neq0   \label{eq:parent}\\  
\textbf{U}_{(l,j)}    &  & \boldsymbol{X}=\textbf{D}_{(l,j)} \text{ for some } l\neq0   \label{eq:parent}\\  
\end{array} \right. \ 
\end{eqnarray}
Note that $\text{Parent}[\cdot]$ does not have definition  for $\textbf{O}_{n}$. We then define $\mathscr{A}_{l,j}$ as all ancestors of qubit $\textbf{U}_{l, j}$ as
\begin{align}
\mathscr{A}_{l,j}=\{\text{Parent}^{\circ t}[\textbf{U}_{l, j}]\big|1\leqslant t\leqslant 3n \}.
\end{align}
Let $\mathscr{A}_{l,j}^{(\text{neighbor})}$ be the set of all the nearest-neighbor qubits of qubits in $\mathscr{A}_{l,j}$, and
\begin{align}
\hat{\mathscr{A}}_{l,j}=\mathscr{A}_{l,j}\cup\mathscr{A}_{l,j}^{(\text{neighbor})}.
\end{align}
As will be demonstrated later, if 
$\boldsymbol{Q}\cap\hat{\mathscr{A}}_{n,j}=\varnothing$, the basis of the final output state with respect to label $j$ is free of errors.

We consider a specific space-time-polarization configuration of error $c\in\mathscr{P}_{\boldsymbol{Q}}$ for some $\boldsymbol{Q}=[Q_1,Q_2,\cdots,Q_M]$. We define the set of survived qubits with respect to $c$ as 
\begin{align}
\mathcal{S}_{\text{surv}}(c)\equiv\{q\in V\big|q\notin Q_{m} \text{ for all } 1\leqslant m\leqslant M\}.
\end{align}
If a qubit is in $\mathcal{S}_{\text{surv}}(c)$, it means that no error has been applied at it during the algorithm. 
We then introduce the set of all \textit{good} branch at the $l$th spatial layer of QRAM as 
\begin{align}
g_l(c)=\left\{j\big|\mathcal{S}_{\text{surv}}(c)\cup \hat{\mathcal{A}}_{l,j}=\varnothing \right\}.
\end{align}
For a lighter notation, we also define 

\begin{align}
\hat{\mathscr{A}}_{j}\equiv\hat{\mathscr{A}}_{n,j},
\end{align}
and
\begin{align}
g(c)\equiv g_n(c).
\end{align}
It turns out that the infidelity is closely related to $g(c)$. In below, we discuss the evolution during fanin phase and fanout phase separately. 

\subsection{Fanin phase}\label{sec:enc}
We assume that before the $l$th step, the quantum state is in the form of 
\begin{align}
|\tilde{\psi}_{l-1}\rangle=E_{l-1}\sum_{j\in g_{l-1}(c)}\psi_{l-1,j}|\mathscr{B}_{l-1,j}\rangle+|\text{garb}_{l-1}\rangle\label{eq:uq}
\end{align}
for some unitary $E_{l-1}$ acting trivially in the good branch, and $|\text{garb}_{l-1}\rangle$ orthogonal to the first term. For a lighter notation, in Eq.~\eqref{eq:uq}, we have neglected the dependency on $c$, and set $|\tilde{\psi}_{l-1}\rangle\equiv|\tilde{\psi}_{l-1}(c)\rangle$, $E_{l-1}\equiv E_{l-1}(c)$ and $|\text{garb}_{l-1}\rangle=|\text{garb}_{l-1}(c)\rangle$.

For $l=0$, Eq.~\eqref{eq:uq} holds because the initial state $|\textbf{U}_{0,0}\rangle=|\mathscr{B}_{0,0}\rangle$ is assumed to be error-free. At the $l$th step, we denote the ideal evolution as $\prod_{j=0}^{2^{l-1}-1}U_{l-1,j}$, with $U_{l-1,j}=\textbf{RT}_{l-1,j}\textbf{CR}_{l-1,j}$. Note that $U_{l-1,j}$ for different $j$ acts on different qubits and do not have overlap, so they commute with each other. Moreover, errors act trivially on qubits in good branches. So we can express the unitary at the $l$th step as 
\begin{align}
\tilde{U}_l=\prod_{j\in g_{l-1}(c)}U_{l,j}\prod_{j\notin g_{l-1}(c)}\tilde{U}_{l,j}.
\end{align}
For $j\notin g_{l-1}(c)$, the unitary $\tilde{U}_{l,j}$ is the noisy implementation of $U_{l,j}$, which acts trivially at good branches. So the quantum state at the $l$th step satisfies 

\begin{align}
|\tilde{\psi}_{l}\rangle=&\tilde{U}_l|\tilde{\psi}_{l-1}\rangle\notag\\
=&\tilde{U}_lE_{l-1}\sum_{j\in g_{l-1}}\psi_{l-1,j}|\mathscr{B}_{l-1,j}\rangle+\tilde{U}_l|\text{garb}_{l-1}\rangle\notag\\
=&\left(\prod_{j\notin g_{l-1}(c)}\tilde{U}_{l,j}\right)E_{l-1}\sum_{j\in g_{l-1}(c)}\tilde{U}_{l,j}\psi_{l-1,j}|\mathscr{B}_{l-1,j}\rangle+\tilde{U}_l|\text{garb}_{l-1}\rangle\notag\\
=&E_{l}\sum_{\lfloor j/2\rfloor\in g_{l-1}(c)}\psi_{l,j}|\mathscr{B}_{l,j}\rangle+\tilde{U}_l|\text{garb}_{l-1}\rangle\label{eq:uq0}\\
=&E_{l}\sum_{j\in g_{l}(c)}\psi_{l,j}|\mathscr{B}_{l,j}\rangle+|\text{garb}_{l}\rangle\label{eq:uq1}.
\end{align}
In Eq.~\eqref{eq:uq0}, we have defined $E_{l}\equiv (\prod_{j\notin g_{l-1}(c)}\tilde{U}_{l,j})E_{l-1}$; in Eq.~\eqref{eq:uq1}, we have defined 
\begin{align}
|\text{garb}_{l}\rangle=E_{l}\sum_{j\in \{j'|\lfloor j'/2\rfloor\in g_{l-1}(c) \& j'\notin g_{l}(c)\}}\psi_{l,j}|\mathscr{B}_{l,j}\rangle+U^{\text{enc}}_l|\text{garb}_{l-1}\rangle.
\end{align}

Accordingly, the final state of the encoding phase is in the form of
\begin{align}
|\tilde{\psi}_n\rangle=\sum_{j\in g(c)}\psi_{j}E|\mathscr{B}_{j}\rangle+|\text{garb}\rangle,
\end{align}
where $E\equiv E_n$, $\mathscr{B}_{j}=\mathscr{B}_{n,j}$, and $|\text{garb}\rangle\equiv|\text{garb}_n\rangle$. Note that unitary $E$ acts trivially at qubits in good branches, and $|\text{garb}\rangle$ is orthogonal to the first term.   

\begin{figure*}[t]
    \centering
          \includegraphics[width=1\columnwidth]{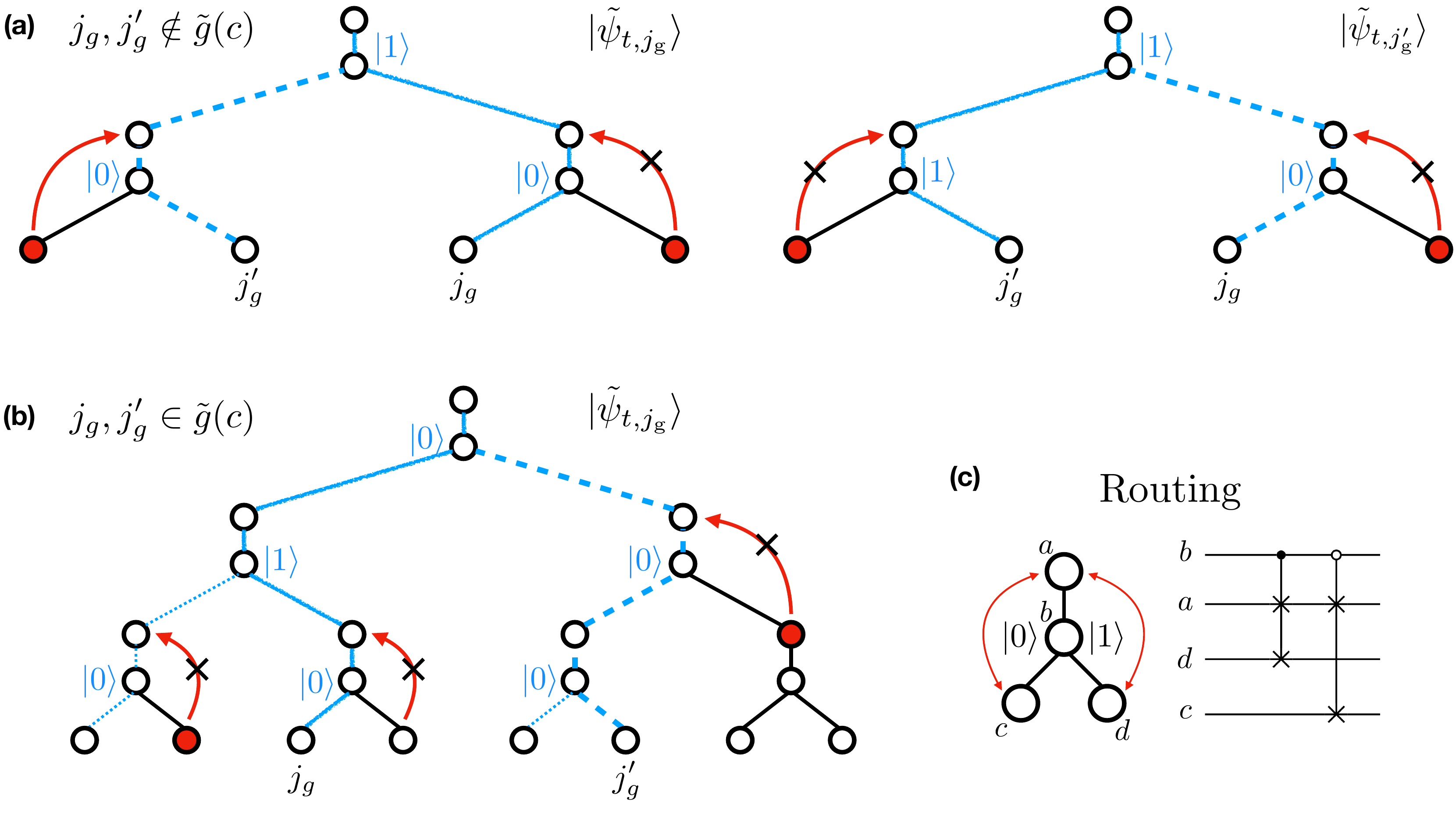}
       \caption{(a) For $j_g,j'_g\in g(c)$ but $j_g,j'_g\notin\tilde{g}(c)$, errors may propagate upward into the good branch from the left hand side. In particular, at the left figure for basis $|\psi_{t,j_g}\rangle$, error propagate into the branch $j'_b$. (b) For basis $|\psi_{t,j_g}\rangle$, routing qubits at all good branches $j'_g\neq j_g$ are always at state $|0\rangle$. With this property, error will never propagate into branches in $\tilde{g}(c)$. (c) Sketch of the routing operations.  } \label{fig:s1}
\end{figure*} 

\subsection{Fanout phase}\label{sec:fan}
We then study the fanout phase. The discussion in the section mainly follows the idea in Sec.V of~\cite{Hann.21}. In the fanout phase, all operations (under a specific error configuration $c$) only transfer a computational basis to another computational basis, up to a phase. So we can always express the quantum state before the $t$th step as
\begin{align}\label{eq:fan0}
|\tilde{\psi}_{t}'\rangle=\sum_{j\in g(c)}\psi_{j}|\psi_{t,j}'\rangle+|\text{garb}'_t\rangle,
\end{align}
where $|\psi_{t,j}'\rangle$ is some computational basis up to a phase, and $|\text{garb}'_t\rangle$ is orthogonal to the first term. Similar to the encoding phase, the expression of states neglect the dependency on $c$. For $t=0$, Eq.~\eqref{eq:fan0} corresponds to $|\psi_{0,j}'\rangle=E|\mathscr{B}_j\rangle$ and $|\text{garb}'_0\rangle=|\text{garb}\rangle$.

At each step, if a routing operation $\textbf{RT}_{l',j'}$ acts nontrivially at a good branch $j\in g(c)$ (this also indicates that it is error-free), then, it can be verified that $\textbf{RT}_{l',j'}$ only swaps $\textbf{U}_{l,j}$ and one of the children qubit of $\textbf{D}_{l,j}$ that is within the branch $j$, while another child qubit of $\textbf{D}_{l,j}$ remains unchanged (see Fig.~\ref{fig:s1}(a)). Moreover, swap gates in the good branch is also error-free. Applying this argument on each elementary routing and swapping operations, the fanout phase performs the basis transformation $E|\mathscr{B}_{j}\rangle\rightarrow|f_j\rangle_{\text{qram}}\otimes|j\rangle_{\text{out}}$ for all $j\in g(c)$. Here, $|f_j\rangle_{\text{qram}}\equiv|f_j(c)\rangle_{\text{qram}}$ is some quantum state of the QRAM. Accordingly,  the final state of the fanout phase, $|\tilde{\psi}'\rangle\equiv|\tilde{\psi}'_{t_{\text{end}}}\rangle$ with $t_{\text{end}}$ the last step, is in the form of 
\begin{align}\label{eq:fan1}
|\tilde{\psi}'\rangle=\sum_{j\in g(c)}\psi_{j}|f_j\rangle_{\text{qram}}\otimes|j\rangle_{\text{out}}+|\text{garb}'\rangle,
\end{align}
for some $|\text{garb}'\rangle$ orthogonal to the first term. 

However, Eq.~\eqref{eq:fan1} is still not sufficient for us to estimate the infidelity. For $j, j'\in g(c)$, we in general have $|f_j\rangle_{\text{qram}}\neq |f_{j'}\rangle_{\text{qram}}$ when $j\neq j'$. After tracing out the QRAM part of Eq.~\eqref{eq:fan1}, the coherence between basis in $g(c)$ may be destroyed. To understand why $|f_j\rangle_{\text{qram}}\neq |f_{j'}\rangle_{\text{qram}}$ (see also Sec.V of~\cite{Hann.21}), we should analyze how error terms propagate from different branches. We first consider basis $|\psi_{t,j_{\text{g}}}\rangle$ with $j_\text{g}\in g(\boldsymbol{Q})$. As shown in left subfigure of Fig.~\ref{fig:s1} (a), suppose an error occurs at the bad branch $j_\text{b}\notin g(\boldsymbol{Q})$, it may propagate into another good branch $j'_{\text{g}}\in g(\boldsymbol{Q})$ ($j'_{\text{g}}\neq j_{\text{g}}$) through a sequence of routing operations (Fig.~\ref{fig:s1}(c)). On the other hand, if we consider the basis $|\psi_{t,j'_{\text{g}}}\rangle$  instead, errors will never propagate into $j'_\text{g}$ (see also right subfigure of Fig.~\ref{fig:s1}). So in general the final state of the QRAM is different for different basis in $g(c)$.

Fortunately, we can identify a large portion of basis in Eq.~\eqref{eq:fan1}, such that errors will still \textit{not} propagate from bad branches to any of the good branches. For these $j$, the final states of QRAM, $|f_j\rangle_{\text{qram}}$, is independent of $j$. To begin with, we notices that in every good branches, error only propagate into it from the right hand side (instead of left hand side). The reason is as follows. Let us consider a basis $|\psi_{t,j_g}\rangle$ with $j_g\in g(c)$. For branch $j_g$, errors will not propagate into it as mentioned previously. For another good branch $j'_{g}\neq j_g$, all routing qubits in it (those in lower sublayers) is at the default state $|0\rangle$. Therefore, swap is only performed between its parent and its right child.  

With the argument above, we suppose $k_1k_2\cdots k_l0\cdots0$ is a good branch, then no errors will ever propagate upward through $\textbf{RT}_{l-1,k}$ (with $k=k_1k_2\cdots k_l$). Therefore, for index $j$, if we have $j_1j_2\cdots j_l0\cdots0\in g(c)$ for all $0\leqslant l\leqslant n-1$, no error will propagate into the branch $j$ from any site. Accordingly, we can define the set of all \textit{error-free} branches
\begin{align}\label{eq:efb}
g'(c)\equiv\{j\in g(c)|\tilde{j}_{l}\in g(c) \text{ for } 0\leqslant l\leqslant n-1\}
\end{align}
where 
\begin{align}
\tilde{j}_{l}\equiv j_1j_2\cdots j_l\underbrace{0\cdots0}_{n-l}.
\end{align}
For the basis $|f_j\rangle_{\text{qram}}\otimes|j\rangle$ of the final state, if $j\in g'(c)$, errors are only applied at good branches. So for all $j\in g'(c)$, their QRAM part is identical, i.e. $|f_{j}\rangle=|f\rangle$ for some quantum state $|f\rangle$. Then, Eq.~\eqref{eq:fan1} can be rewritten as 
\begin{align}\label{eq:fan2}
|\tilde{\psi}'\rangle=\sum_{j\in g'(c)}\psi_{j}|f\rangle_{\text{qram}}\otimes|j\rangle_{\text{out}}+|\widetilde{\text{garb}}'\rangle.
\end{align}
Note that $|f\rangle$ is independent of $j$, but still depends on $c$. 

\subsection{State preparation infidelity}\label{sec:spi}
In Sec.~\ref{sec:fan}, the final output state is $|\tilde{\psi}'\rangle$. Comparing Eq.~\eqref{eq:fan2} to Eq.~\eqref{eq:roc}, we have 
\begin{align}\label{eq:fan3}
\tilde{\rho}_{\text{out}}(c)=\text{Tr}_{\text{qram}}\left[|\tilde{\psi}'\rangle\langle\tilde{\psi}'|\right].
\end{align}
We define $|\psi'\rangle\equiv\sum_{j=1}^{N-1}\psi_{j}|f\rangle_{\text{qram}}\otimes|j\rangle_{\text{out}}$. The fidelity between $|\psi'\rangle$ and Eq.~\eqref{eq:fan2} is  
\begin{align}
\text{Fid}\left(|\psi'\rangle\langle\psi'|,|\tilde{\psi}'\rangle\langle\tilde{\psi}'|\right)=\sum_{j\in g'(c)}|\psi_j|^{2}\equiv \Lambda'(c).
\end{align}
Here, $\Lambda'(c)$ highlights that it is depends on $c$.
 Because fidelity is non-decreasing under partial trace, we have 
\begin{align}
\text{Fid}\left(\text{Tr}_{\text{qram}}\left[|\psi'\rangle\langle\psi'|\right],\tilde{\rho}(c)\right)\geqslant \Lambda'(c).
\end{align}
Moreover, it can be verified that $\text{Tr}_{\text{qram}}\left[|\psi'\rangle\langle\psi'|\right]=\rho_{\text{id}}$. So 
\begin{align}
\text{Fid}\left(\rho_{\text{id}},\tilde{\rho}(c)\right)\geqslant\Lambda'(c).
\end{align}
Combining with Eq.~\eqref{eq:cov}, the total state preparation infidelity satisifes
\begin{align}\label{eq:fid2}
F\geqslant\mathbb{E}[\Lambda'(c)],
\end{align}
where $\mathbb{E}[\Lambda'(c)]$ represents the expectation value of $\Lambda(c)$ when sampling $c$ according to $p_c$.

We now estimate $\mathbb{E}[\Lambda'(c)]$. Let $\mathscr{J}=\{0,1,\cdots, N-1\}$ be all indexes, and $\text{Power}(\mathscr{J})$ be the set of all subset of $\mathscr{J}$. By definition, we have
\begin{align}\label{eq:JJ}
\mathbb{E}[\Lambda'(c)]&=\sum_{J\in\text{Power}(\mathscr{J})}\sum_{j\in J}|\psi_{j}|^2 \times\text{Pr}[J_1\in g'(c)]\text{Pr}[J_2\in g'(c)|J_1\in g'(c)]\text{Pr}[J_3\in g'(c)|J_1,\mathscr{J}_2\in g'(c)]\cdots.
\end{align}
In Eq.~\eqref{eq:JJ}, $J_1,J_2,\cdots$ are elements of $J$ arranged in arbitrary order. Note that different branches may have overlap, and we always have 
\begin{align}\label{eq:JJ2}
&\text{Pr}[J_2\in g'(c)|J_1\in g'(c)]\geqslant\text{Pr}[J_2\in g'(c)],\\
&\text{Pr}[J_3\in g'(c)|J_1,\mathscr{J}_2\in g'(c)]\geqslant\text{Pr}[J_3\in g'(c)],
\end{align}
and so on. Therefore, we have
\begin{align}
\mathbb{E}[\Lambda'(c)]\geqslant&\sum_{J\in\text{Power}(\mathscr{J})}\sum_{j\in J}|\psi_j|^2 \times\text{Pr}[J_1\in g'(c)][J_2\in g'(c)][J_3\in g'(c)]\cdots\label{eq:exp1}\\
=&\sum_{j=0}^{N-1}|\psi_{j}|^2\text{Pr}[j\in g'(c)]\label{eq:exp2}\\
=&\text{Pr}[j\in g'(c)]\label{eq:exp3}
\end{align}
Eq.~\eqref{eq:exp2} is because the right hand side of Eq.~\eqref{eq:exp1} corresponds to a summation of multiple variables sampled independently. Eq.~\eqref{eq:exp3} is because of the normalization of $\psi_j$, and the probability is independent of $j$. By definition in Eq.~\eqref{eq:efb}, $j$ is an error-free branch in $g'(c)$, if and only if all qubits in $\tilde{j}_l$ (for all $0\leqslant l\leqslant n-1$) are free of error at all time. There are at most $O(n^2)$ of these qubits. For each individual qubit, the probability that it is error free at all time is $(1-\varepsilon)^{O(n)}$, because the algorithm has totally $O(n)$ steps. Therefore, with probability $((1-\varepsilon)^{O(n)})^{O(n^2)}=(1-\varepsilon)^{O(n^3)}$, $j$ is a good branch. By Bernoulli inequality, we have $\text{Pr}[j\in g'(c)]\geqslant1-A\varepsilon n^3$ for some constant $A$. So we have
 \begin{align}
\mathbb{E}[\Lambda'(c)]\geqslant (1-A\varepsilon n^3).
\end{align}
Combining with Eq.~\eqref{eq:fid2}, we have 
 \begin{align}
1-F\leqslant A\varepsilon n^3.
\end{align}

\section{Proof of the robustness for 3-qubit-per-node protocol}\label{sec:3rb}
\subsection{Robustness analysis}

\begin{figure*}[t]
    \centering
          \includegraphics[width=1\columnwidth]{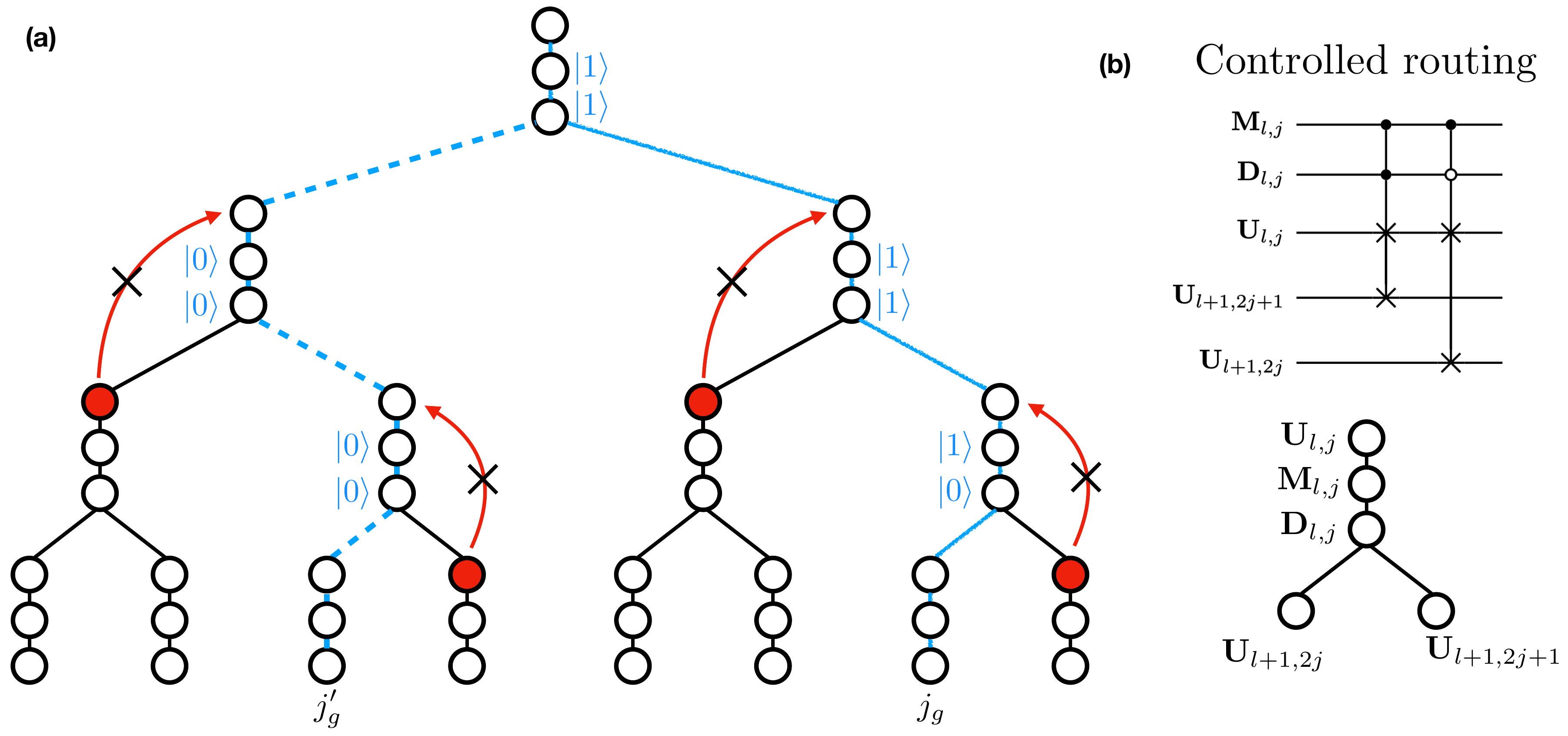}
       \caption{(a) For both $j_g\in g(c)$ and $j'_g\in g(c)$, errors never propagate into the good branches (blue color), because all controlled qubits $\textbf{M}_{l,j}$ in good branches are error free.  (c) Sketch of the controlled routing operations.} \label{fig:s2}
\end{figure*}

With an abuse of notation, we define \textit{good} index and relevant terminologies here in a similar way to the 2-qubit-per-node protocol. Let $\mathscr{A}_{j}$ be all ancestors of $\textbf{U}_{n,j}$ in both QRAM and output register. We also define $\hat{\mathscr{A}}_j$ as the intersection of $\mathscr{A}_{j}$ and its nearest neighbour. Similar to the 2-qubit-per-node protocol, for a specific space-time-polarization configuration of error $c$, we define $g(c)$ as set of all \textit{good} index $j$, such that all qubits in $\hat{\mathscr{A}}_j$ are free of errors at all time. 

With the same argument to the 2-qubit-per-node protocol, the final output state of can be expressed as 

\begin{align}\label{eq:fan1}
|\tilde{\psi}'\rangle=\sum_{j\in g(c)}\psi_{j}|f_j\rangle_{\text{qram}}\otimes|j\rangle_{\text{out}}+|\text{garb}'\rangle
\end{align}
for some garbage state that is orthogonal to the first term. Yet, the main difference is that in the 3-qubit-per-node protocol here, errors will never propagate into the good branches $j\in g(c)$ (as oppose to $j\in g'(c)$ in the 2-qubit-per-node protocol). The reason is as follows (see also Fig.~\ref{fig:s2}). During the fanout process, we suppose the quantum state at a certain step $t$ is 
\begin{align}\label{eq:fan1}
|\tilde{\psi}'_{t}\rangle=\sum_{j\in g(c)}\psi_{j}|\psi'_{t,j}\rangle_{\text{out}}+|\text{garb}_t'\rangle.
\end{align}
We now consider basis $|\psi_{t,j_b}'\rangle$ for some $j_g\in g(c)$. During controlled routing operations, errors will not propagate into the branch $j_g$, because the controlled and routing qubits are at correct state. We now consider other good branch $j_g'\in g(c)$ that $j_g'\neq j_g$. All of their control qubits in the middle sublayers are free of errors, and hence at state $|0\rangle$. Therefore, all corresponding routing operations does not perform any swapping, and errors will not propagate from bad branch to the branch $j_g'$. 

As a result, errors perform trivially at all good branches, so for all $j\in g(c)$, we have $|f_j\rangle_{\text{qram}}=|f\rangle_{\text{qram}}$ for some computational basis $f$ independent of $j$. Let $\Lambda=\sum_{j\in g(c)}|\psi_j|^2$, with the same argument for obtaining Eq.~\eqref{eq:fid2} in Sec.~\ref{sec:spi}, we have $F\geqslant \mathbb{E}[\Lambda]$. Similar to Eq.~\eqref{eq:exp3}, we also have 
\begin{align}
\mathbb{E}[\Lambda]\geqslant \text{Pr}[j\in g(c)].
\end{align}
Because $\mathscr{A}_j=O(n)$ and the algorithm has runtime $O(n)$, we have $\text{Pr}[j\in g(c)]\geqslant(1-\varepsilon)^{O(n)\times O(n)}\geqslant1-A\varepsilon n^2$ for some constant $A$. Therefore, the total infidelity satisfies  
\begin{align}
1-F\leqslant A\varepsilon n^2.
\end{align}

\end{appendices}

\end{document}